\documentclass[apj]{emulateapj}
\usepackage{graphicx}

\def\HI{{\sc Hi}}
\def\HII{{\sc Hii}}
\def\Msol{M_\odot}
\shorttitle{W49B and its environment}
\shortauthors{Zhu, Tian and Zuo}

\begin{document}

\title{supernova remnant W49B and its environment}

\author{H. Zhu\altaffilmark{1}, W. W. Tian\altaffilmark{1,2}, P. Zuo\altaffilmark{1}}
\affil{$^1$Key Laboratory of Optical Astronomy, National Astronomical Observatories, Chinese Academy of Sciences,\\
Beijing 100012, China; zhuhui@bao.ac.cn and tww@bao.ac.cn}
\affil{$^2$Department of Physics \& Astronomy, University of Calgary, Calgary, Alberta T2N 1N4, Canada}

\begin{abstract}
We study Gamma-ray supernova remnant W49B and its environment using recent radio and infrared data. {\it Spitzer} IRS low resolution data of W49B shows shocked excitation lines of H$_{2}$ (0,0) S(0)-S(7) from the SNR-molecular cloud interaction. The H$_2$ gas is composed of two components with temperature of $\sim$260 K and $\sim$1060 K respectively. Various spectral lines from atomic and ionic particles are detected towards W49B. We suggest the ionic phase has an electron density of $\sim$500 cm${}^{-3}$ and a temperature of $\sim$${10^4}$ K by the spectral line diagnoses. The mid- and far-infrared data from {\it MSX}, {\it Spitzer} and {\it Herschel} reveals a 151 $\pm$ 20 K hot dust component with a mass of 7.5 $\pm$ 6.6 $\times$ ${10}^{-4} {\Msol}$ and a 45 $\pm$ 4 K warm dust component with a mass of 6.4 $\pm$ 3.2 ${\Msol}$. The hot dust is likely from materials swept up by the shock of W49B. The warm dust may possibly originate from the evaporation of clouds interacting with W49B. We build the HI absorption spectra of W49B and nearby four {\HII} regions (W49A, G42.90+0.58, G42.43-0.26 and G43.19-0.53), and study the relation between W49B and the surrounding molecular clouds by employing the 2.12 $\mu$m infrared and CO data. We therefore obtain a kinematic distance of $\sim$10 kpc for W49B and suggest that the remnant is likely associated with the CO cloud at about 40 km s$^{-1}$. \\
\end{abstract}


\section{Introduction}
\noindent Massive stars ($\geq 8 {M_\odot}$) evolve quickly. After they end with core-collapse supernova explosion and form supernova remnants (SNRs), their mother molecular clouds should be still nearby. It is expected that the mother clouds will shape the SNRs significantly and SNRs also remold the surrounding medium, e.g. ejecting newly formed dust to the medium, changing the ingredient of dust swept up by the shock, triggering molecule cloud core collapse and producing cosmic rays.\\

W49B is a mixed-morphology (MM) SNR and one of the brightest sources in the Galaxy at 1 GHz (\citealt{keo07}, \citealt*{mof94}). Multifrequency radio observations of W49B show a 4${'}$ box or barrel like structure with amount of distorted filaments. W49B has an integrated spectral index of about -0.48. Polarization is detected at 6 cm but with very low mean polarized fraction of $0.44\% \pm 0.06\%$. \citet*{mof94} explained the low fraction as depolarization caused by tangled magnetic fields on the shell or Faraday depolarization within the remnant. The 1.64 $\mu$m [Fe II] image reveals similar barrel like structure as the radio displayed. The shocked molecular hydrogens traced by the H$_2$ (1,0) S(1) line at 2.12 $\mu$m are found nearly surrounding W49B with strong emission at the east, south and west boundaries. \citet{keo07} suggested these as evidences of W49B born in a wind-blown bubble inside a dense molecular cloud.\\

Early X-ray observations carried out by {\it Einstein}, {\it EXOSAT} and {\it ASCA} (\citealt{pye84}; \citealt{smi85}; \citealt{fuj95}; \citealt{hwa00}) show the X-ray emission is thermal and ejecta dominated. Based on these features, the age of W49B is estimated from 1000 yr to 4000 yr. {\it ASCA} observation discloses that the Fe emissions trend to rise in the inner part of the SNR while the emission of Si, S extents to the outer region. It also provides evidence of overabundances of Si, S, Ar, Ca and Fe through fitting the global spectrum. Thanks to the high spacial resolution and sensitivity of {\it XMM-Newton}, {\it Chandra} and {\it Suzaku}, more striking secrets about W49B were exposed (\citealt{mic06,mic10}, \citealt{lop09,lop13a,lop13b}, \citealt{oza09}, \citealt{yan13}). First, W49B has the most elliptical and elongated morphology in young, X-ray bright SNRs. Second, elemental segregation is confirmed with the Si, S, Ar, Ca showing more extended homogeneous structure but the iron nearly missing from the west. Third, silicon burning products, Cr and Mn, are detected in W49B's spectrum. Finally, W49B shows clear overionization feature which may be caused by rapid cooling. Other SNRs with overionized plasmas usually have ages older than $\sim$4000 yr. Since W49B is young, this indicates that overionization can occur in the SNR's early evolution stage.\\

GeV and TeV $\gamma$-rays have been detected just coincident with the radio brightest part of W49B (\citealt{abd10}, \citealt{bru11}). The observed $\gamma$-ray photon spectrum is very steep. Its spectral index is at least 0.5 higher than the electron spectral index. This favors a hadronic origin for the $\gamma$-ray radiation (\citealt{man11}). Other evidences supporting the hadronic model include: (1) the observed high GeV $\gamma$-ray luminosity and (2) dense molecular clouds interacting with W49B (\citealt{bru11}). More detail information about the ambient of W49B is needed for further study of the $\gamma$-ray emission.\\

To explain the characters of W49B, two scenarios are proposed: one is a jet driven bipolar explosion of a massive star, another is a normal spherical supernova exploded in inhomogeneous medium. Currently there are more evidences supporting the first model (\citealt{lop13a}): (1) near infrared and radio images show a barrel like morphology which is common in bubble blown by massive star; (2) X-ray data reveals the jet like elongated structure; (3) the abundance of Si, S, Ar and Ca (relative to Fe) favors an aspherical explosion by a 25 ${\Msol}$ progenitor; (4) no neutron star or pulsar is detected in W49B. The normal spherical supernova model can explain most of the observed traits. However, this model is not able to produce the observed element abundance and the ejecta mass is also too big to be explained easily (\citealt{mic10}). Therefore the picture that W49B originates from a jet driven bipolar explosion of a 25 ${\Msol}$ massive star with its morphology shaped by the interstellar medium is more favorite. But how the environment effects the morphology of W49B is still little known (\citealt{zho11}).\\

Distance is another confusion for W49B. {\HI} absorption spectra toward W49B and its nearby {\HII} region W49A which has a distance of 11.4 kpc (\citealt*{gwi92}) have been given by \citet{rad72}, \citet*{loc78} and \citet{bro01}. Compared with W49A, W49B is lack of {\HI} absorption at velocities of $\sim$10 km s${}^{-1}$ and $\sim$55 km s${}^{-1}$. The absence of absorption at 55km s${}^{-1}$ hints W49B is $\sim$3 kpc closer than W49A (\citealt{rad72}). \citet{bro01} found the ${N_{HI}}/{T_s}$ image toward W49A and W49B have obvious change on size scales of about 1${'}$. They thought the differences in the absorption spectra of W49A and W49B could be explained without assuming W49B is closer, but with scenarios involving differences on kinematics, distribution and temperature. They suggested W49B might have the same distance as W49A based on the possible interaction between the SNR and the {\HI} cloud at $\sim$5 km s$^{-1}$. Recently, \citet{che14} noticed that $\textrm{CO} (J=2-1)$ emission map at 39-40 km s$^{-1}$ for W49B show possible cavity structure, but no robust kinematic evidence has been obtained yet. The authors suggested if this SNR-molecular cloud association is true, the kinematic distance to W49B is 9.3 kpc.\\

To understand W49B and its environment better, we perform a radio and infrared study toward this SNR. In section 2, we give a description about the data. The results are presented in section 3. A discussion about the properties of W49B and its environment, such as the distance, shock and dust, is in section 4. Section 5 is a summary.\\

\section{Data and data reduction}

\subsection{Radio data}

\noindent The 1420 MHz radio continuum and {\HI} emission data come from the Very Large Array Galactic Plane Survey \citep[VGPS,][]{sti06}. The continuum data has a spatial resolution of $1{'}$ and a noise of 0.3 K. The {\HI} spectral line images have a resolution of $1{'} \times 1{'} \times$ 1.56 km s${}^{-1}$ and an noise of 2 K per channel. We also use the 20 cm continuum data from the Multi-Array Galactic Plane Imaging Survey which have resolution of $\sim$6${''}$ (MAGPIS, \citealt{hel06}). The ${}^{13}\textrm{CO} (J=1-0)$ spectral line data is from the BU-FCRAO Galactic Ring Survey \citep{jac06}. The data has an angular and spectral resolution of $46{''}$ and 0.21 km s${}^{-1}$ with noise of about 0.13 K, respectively. We get the $\textrm{CO} (J=3-2)$ data from the first release of {\it JCMT} CO High-Resolution Survey (\citealt*{dem13}). The data is smoothed to a velocity resolution of 1 km s${}^{-1}$ and a spatial resolution of $16.6{''}$. The noise in the region of W49B is about 0.44 K.\\

\subsection{Infrared data}

\noindent The {\it Spitzer} Infrared Spectrograph (IRS, \citealt{hou04}) observation of W49B is taken on 24 April 2005 using the staring mode (Program ID: 3483, PI: Rho, J.). The short-low slit (SL, 5.2-14.5 $\mu$m) and long-low slit (LL, 14.0-38.0 $\mu$m) are used in the observation. The bad and rogue pixels of the co-added BCD frames are corrected with the IRSCLEAN tool provided by the Spitzer Science Center. Since W49B is extended, only different orders are used to remove the unrelated diffuse emission in the direction of the SNR. After background subtraction, an optimal spectrum is extracted by the SPICE software. The spectrum extraction region is selected to meet criterion of the brightest overlapped region between SL and LL. Error in the final extraction is of the order of 10$\%$ which are mainly caused by the background subtraction (\citealt{and11}). \\

The {\it Herschel} observation are taken on 24 October 2010 as part of the {\it Herschel} Infrared Galactic Plane Survey (Hi-GAL). The PACS (\citealt{pog10}) and SPIRE (\citealt{gri09}) are used under the parallel mode. The covered bands are 70 $\mu$m, 160$ \mu$m, 250 $\mu$m, 350 $\mu$m and 500 $\mu$m. Each bands have a measured beam of $\sim$9.7${''}$ $\times$ $\sim$10.7${''}$, $\sim$13.2${''}$ $\times$ $\sim$13.9${''}$, $\sim$22.8${''}$ $\times$ $\sim$23.9${''}$, $\sim$29.3${''}$ $\times$ $\sim$31.3${''}$ and $\sim$41.1${''}$ $\times$ $\sim$43.8${''}$ respectively (\citealt{tra11}). The PACS bands (70 $\mu$m and 160 $\mu$m) have been processed using the standard pipeline. Since the SPIRE data has a problem of stripe, we re-process them by combining the scripts `Photometer Map Merging' with `Baseline Removal and Destriper' to merge the two direction scan data to a single map. We also remove the SPIRE observation baseline and do a correction for the relative gain of bolometer. The process uses scripts within the software package {\it Herschel} Interactive Processing Environment (HIPE, version 11.0.1) and calibration tree 11.0, which was the most up-to-date version at the time.\\

The 2.12 $\mu$m data is from `UKIRT Widefield Infrared Survey' for H$_2$ (\citealt{fro11}). Other infrared continuum image of {\it WISE} 12 $\mu$m (\citealt{wri10}), {\it Spitzer} 24 $\mu$m (\citealt{car09}), {\it MSX} 12.1 $\mu$m, 14.7 $\mu$m and 21.3 $\mu$m (\citealt{pri01}) are collected directly from NASA/IPAC Infrared Science Archive.\\

\begin{figure*}[!htpb]
\centerline{\includegraphics[width=0.49\textwidth, angle=0]{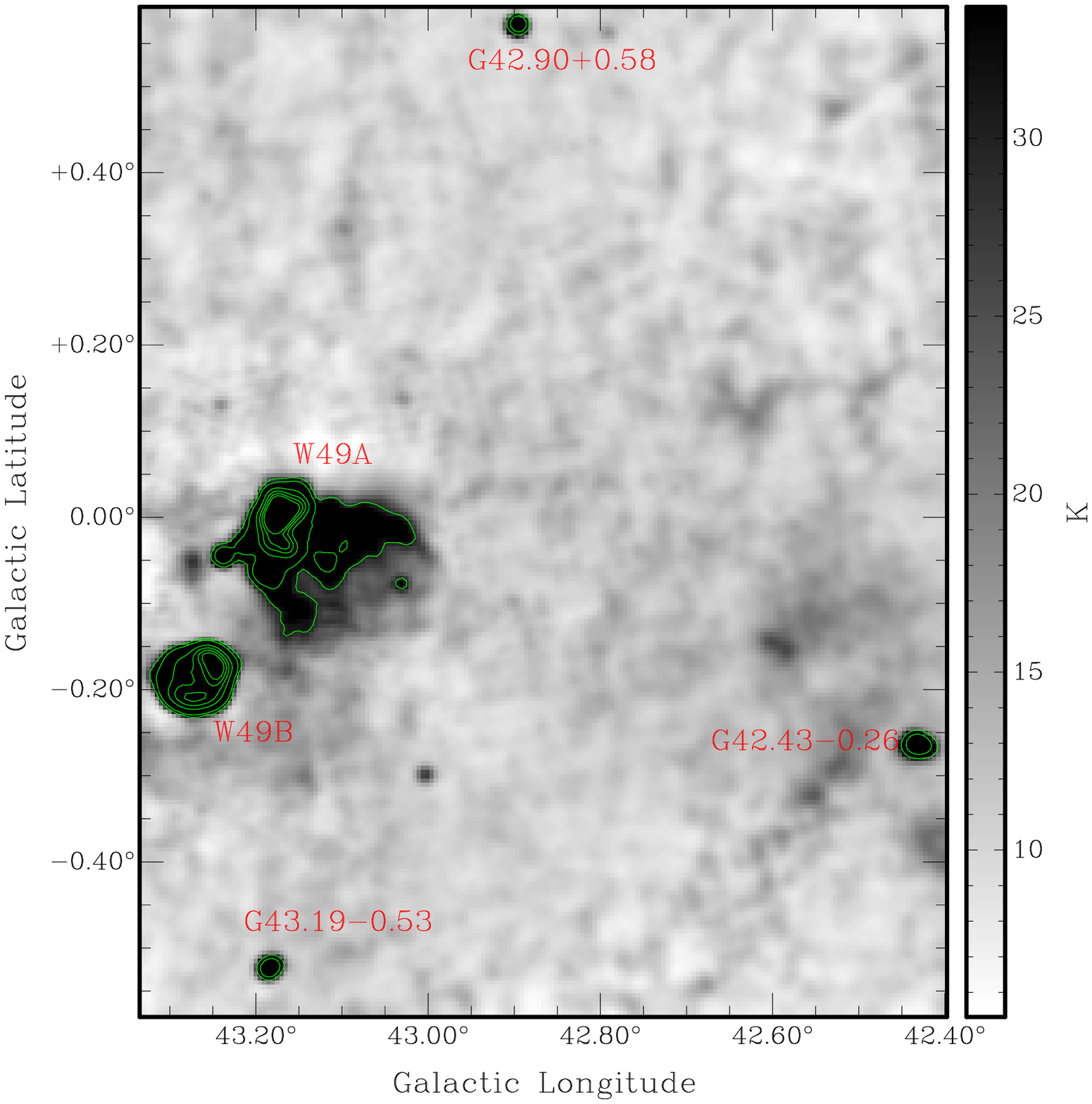}\includegraphics[width=0.49\textwidth, angle=0]{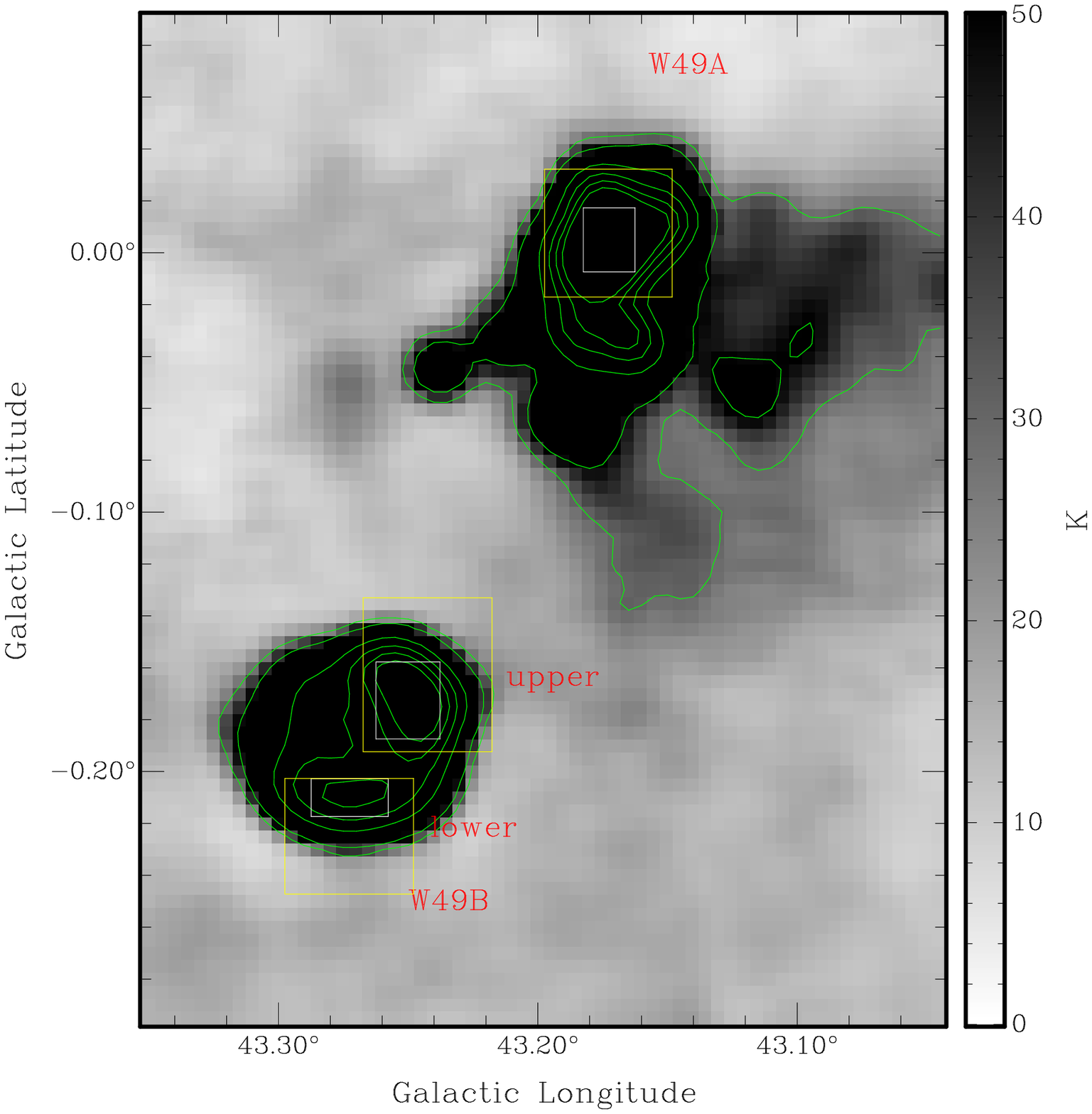}}
\caption{Left: The VGPS 1420 MHz continuum image around W49B with contour levels of 30, 50, 150, 250, 350 and 450 K. Right: White boxes are the source-on regions to extract the absorption spectra, and the regions between white and yellow boxes are used to subtract the background.}
\label{fig:1}
\end{figure*}

\section{Results}

\subsection{{\HI} absorption spectra}

\noindent The left panel of Figure \ref{fig:1} displays the 1420 MHz continuum image around W49B. We build the {\HI} absorption spectrum by the revised methods presented by \citet{tia07} and \citet*{tia08}. The methods minimize the possibility of a false absorption spectrum due to the potential {\HI} distribution difference in two lines of sight. In the right panel, the marked white boxes are used to extract the spectra. Regions between the white boxes and yellow boxes are the background. For W49B, we select two bright regions to analyze its {\HI} absorption spectrum.\\

\begin{figure*}[!htpb]
\centerline{\includegraphics[width=0.5\textwidth, angle=0]{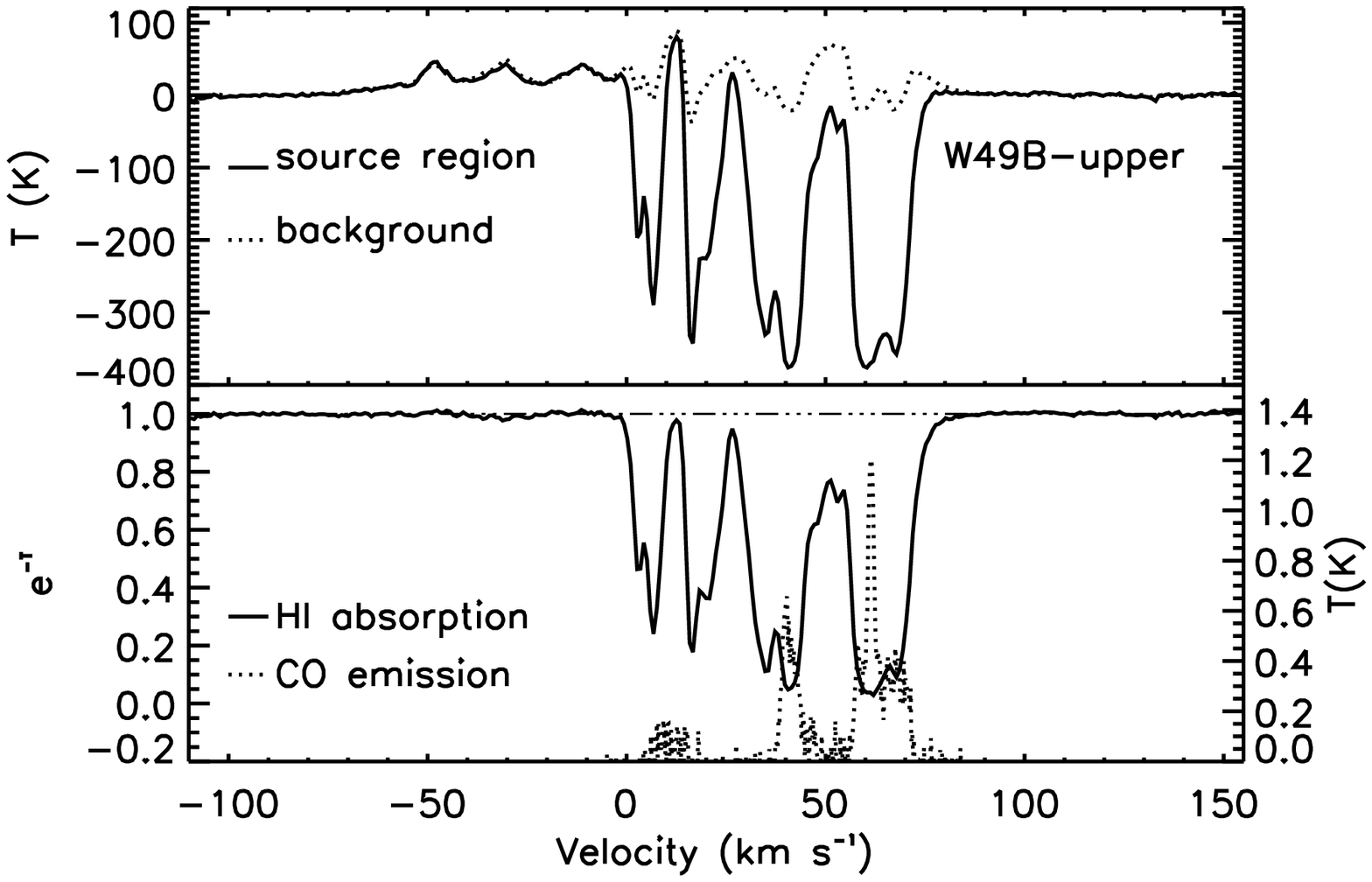}\includegraphics[width=0.5\textwidth, angle=0]{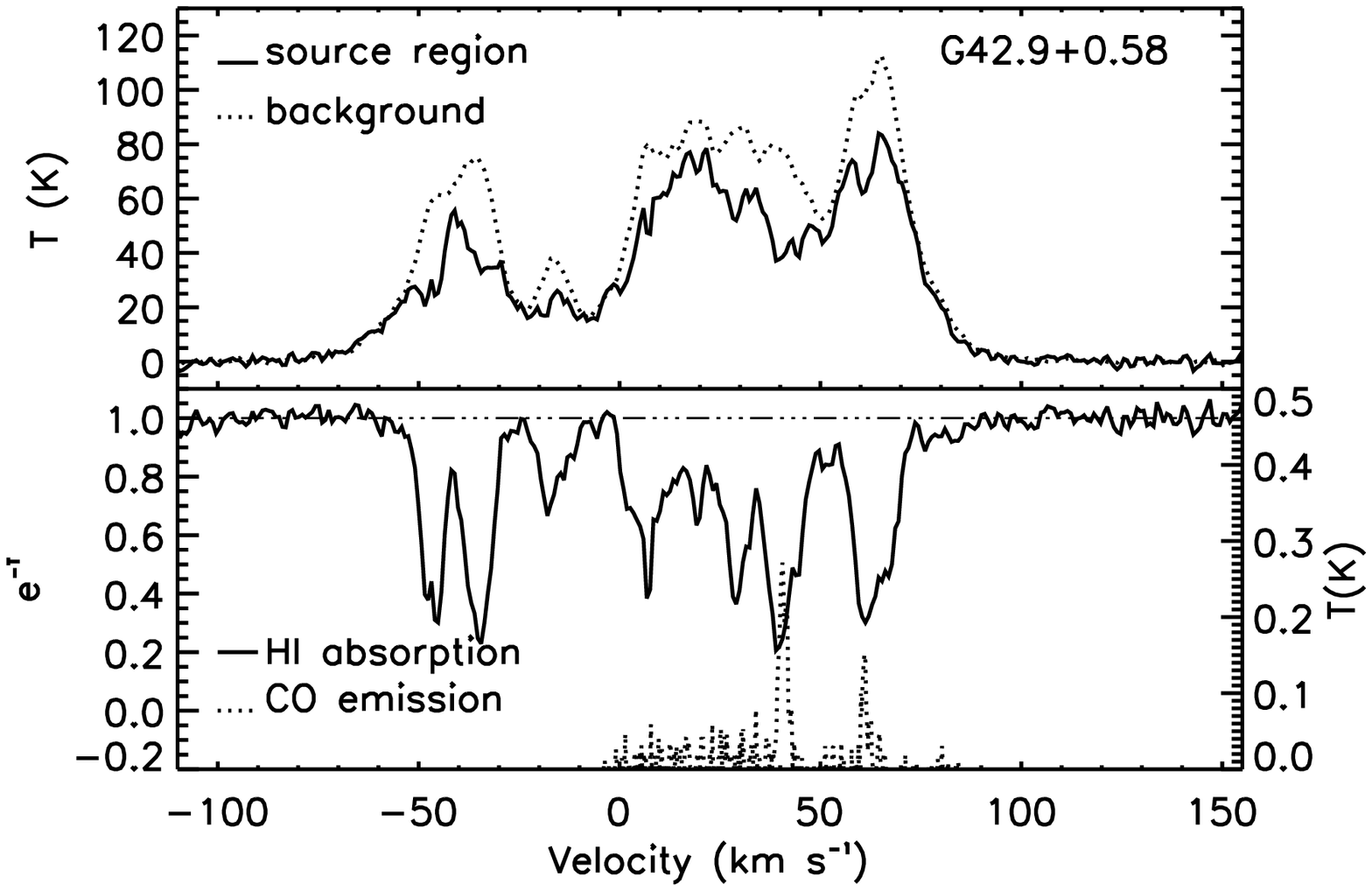}}
\centerline{\includegraphics[width=0.5\textwidth, angle=0]{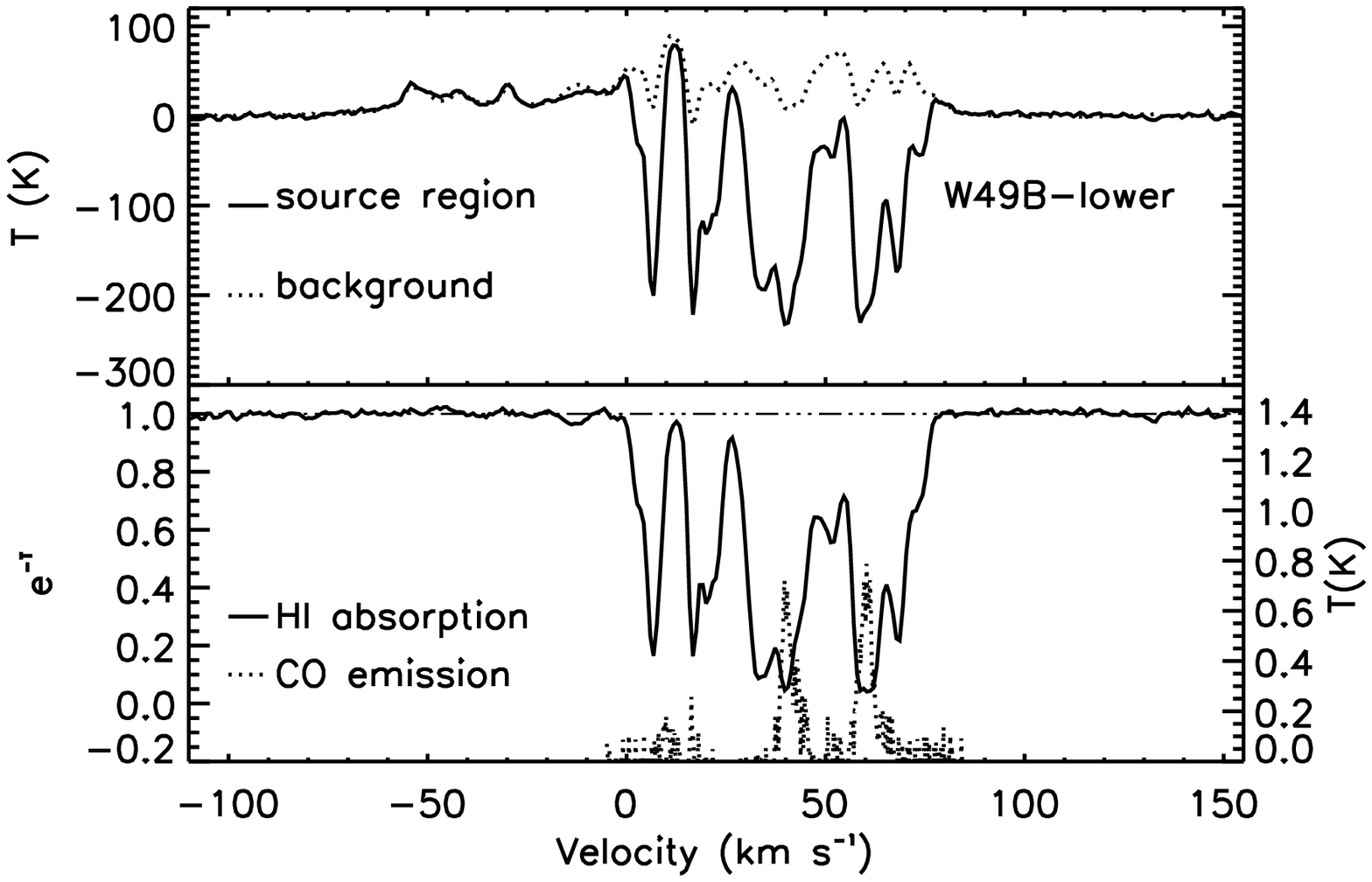}\includegraphics[width=0.5\textwidth, angle=0]{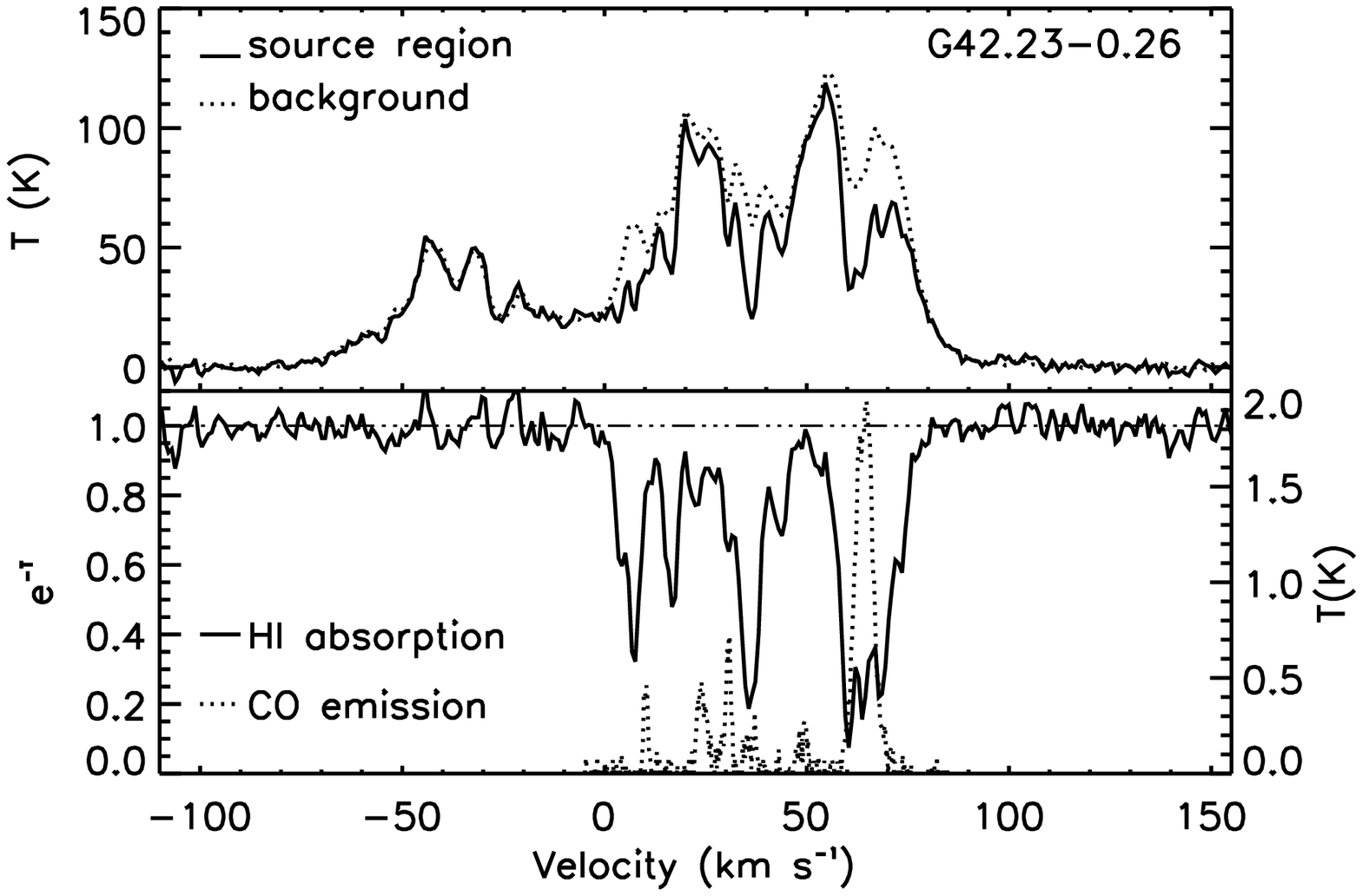}}
\centerline{\includegraphics[width=0.5\textwidth, angle=0]{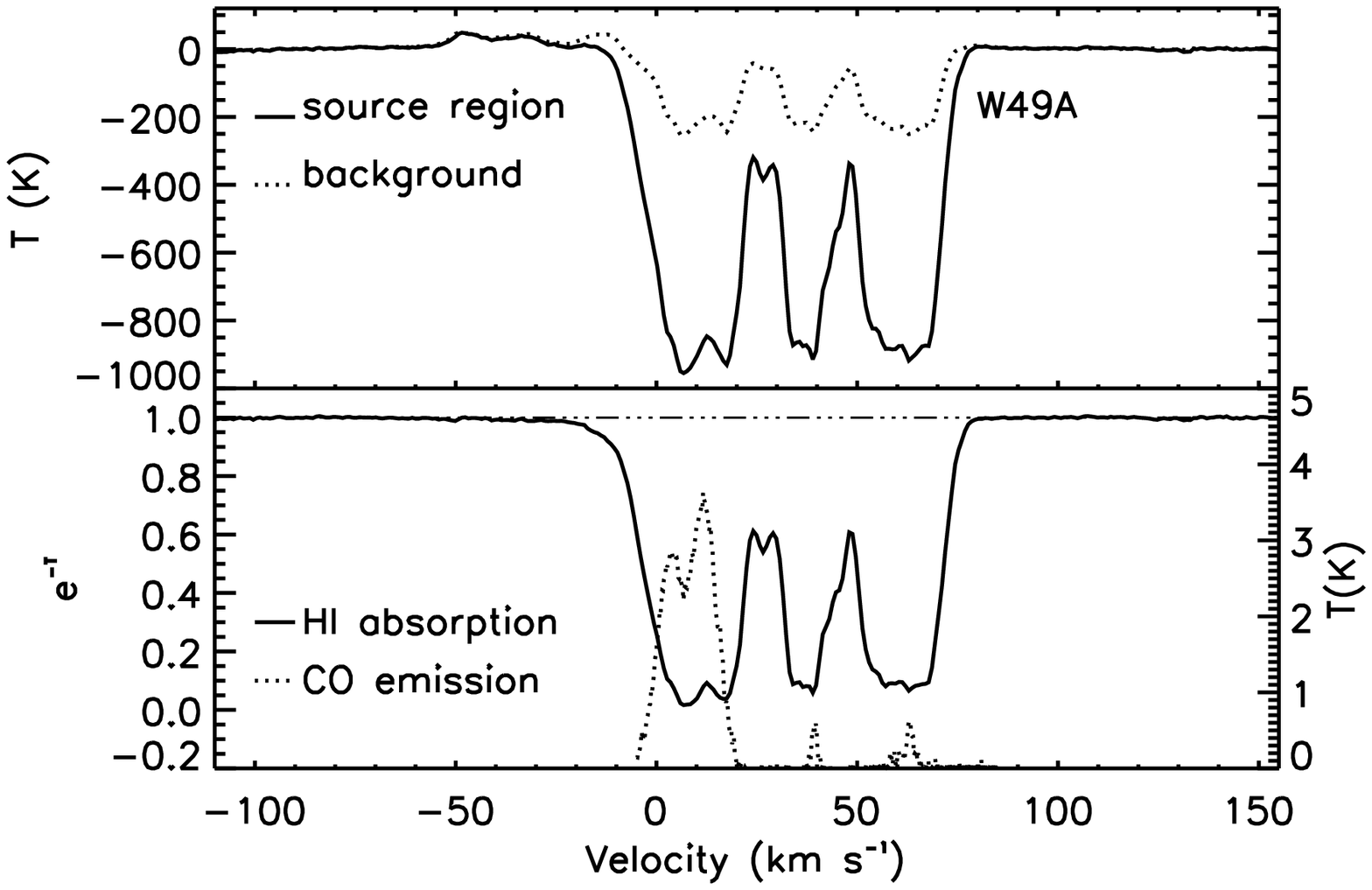}\includegraphics[width=0.5\textwidth, angle=0]{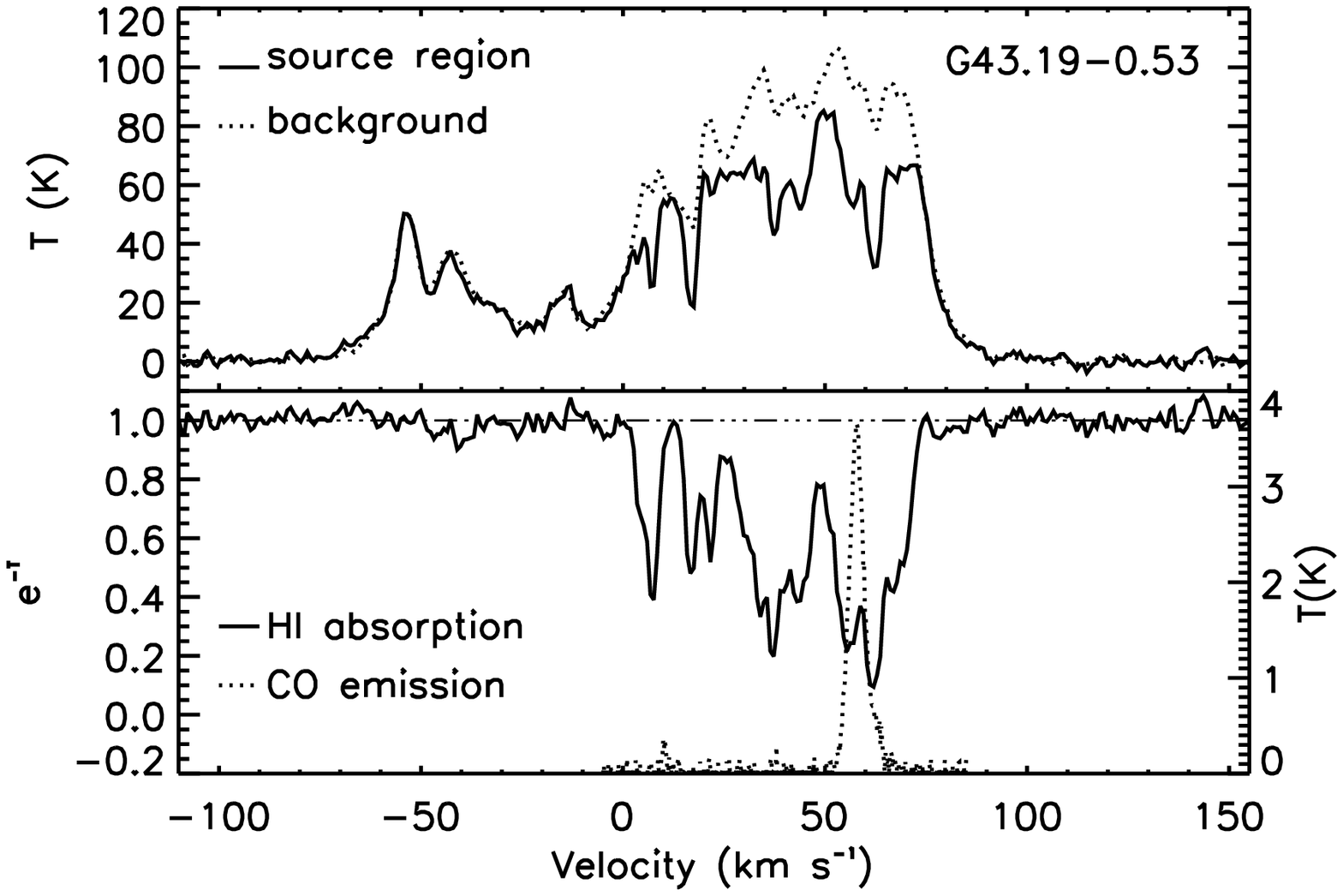}}\caption{{\HI} absorption spectra of W49B, W49A and three nearby {\HII} regions, G42.9+0.58, G42.43-0.26 and G43.19-0.53. In the lower part of each panel, the left Y axis means {\HI} optical depth in the form of $e^{-\tau}$ and the right Y axis represents the brightness temperature of ${}^{13}$CO.}
\label{fig:2}
\end{figure*}

Figure \ref{fig:2} shows the {\HI} absorption spectra of W49B, W49A and three nearby {\HII} regions. The ${}^{13}\textrm{CO} (J=1-0)$ spectral lines in each panel are extracted from the same regions as {\HI} spectra. There are three main {\HI} absorption features in the spectra which are at $\sim$10 km s${}^{-1}$, the $\sim$40 km s${}^{-1}$ and the $\sim$60 km s${}^{-1}$ respectively. For {\HII} regions G42.43-0.26 and G43.19-0.53, radio recombination lines are detected at 62.7 km s${}^{-1}$ and 55.0 km s${}^{-1}$ (\citealt*{loc89}). Since {\HI} absorptions appear up to $\sim$70 km s${}^{-1}$, the two {\HII} regions are located at the far side distances, i.e. $\sim$7.8 kpc and $\sim$8.4 kpc respectively (using the Galactic rotation curve model with ${R}_{\odot} =$ 8.5 kpc and ${V}_{\odot} =$ 220 km s${}^{-1}$). Therefore the $\sim$10 km s${}^{-1}$ absorption of G42.43-0.26 and G43.19-0.53 likely originates from local {\HI} clouds. {\HII} region G42.90+0.58 (\citealt*{woo89}) is located outside the solar circle because of its absorption at negative velocity. The trigonometric parallax measurement (\citealt*{gwi92}) gives W49A a distance of 11.4 kpc which is consistent with the far side kinematic distance of $\sim$10 km s${}^{-1}$. Compared with the absorption spectra of G42.43-0.26 and G43.19-0.53, stronger absorption at $\sim$10 km s${}^{-1}$ is detected towards both W49A and G42.90+0.58. This is because the absorption at 10 km s${}^{-1}$ of W49A and G42.90+0.58 arises from both local and Perseus spiral arm.\\

\begin{figure*}[!htpb]
\centerline{\includegraphics[width=0.49\textwidth, angle=0]{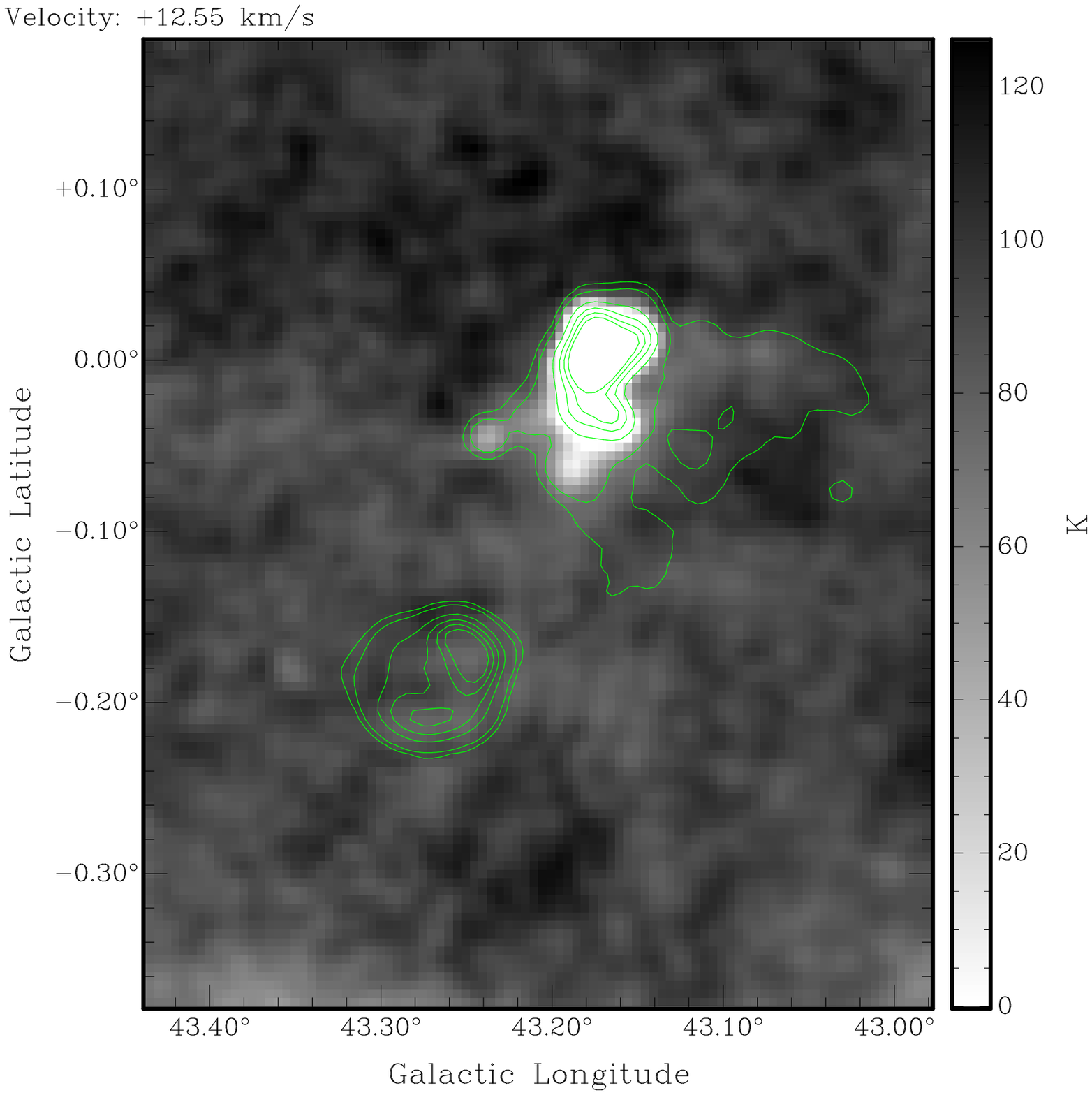}\hfill\includegraphics[width=0.49\textwidth, angle=0]{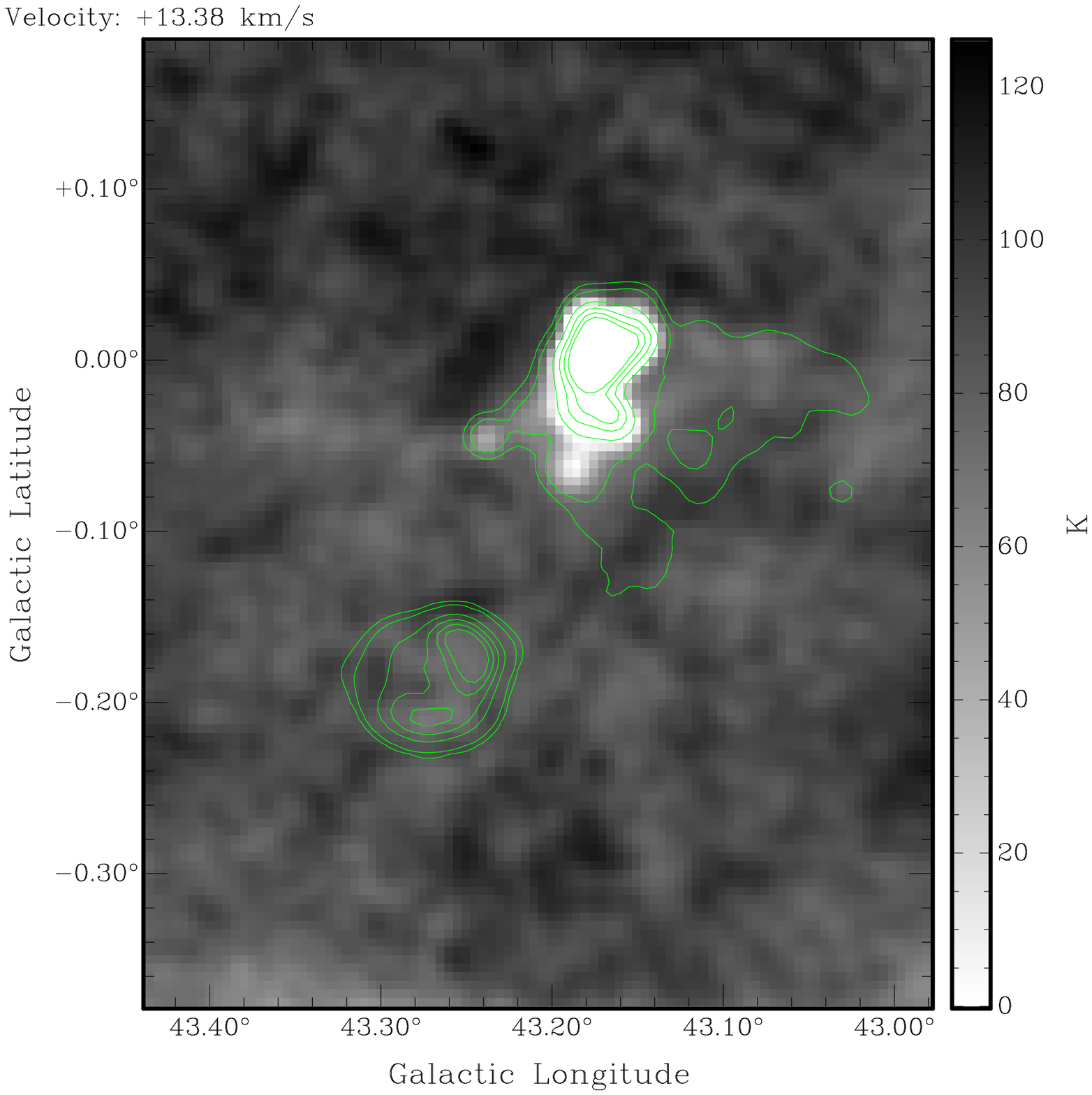}}
\caption{{\HI} channel maps at velocities of 12.55 km s${}^{-1}$ (Left) and 13.38 km s${}^{-1}$ (Right) with the contour same as Figure \ref{fig:1}.}
\label{fig:3}
\end{figure*}

The absorption spectrum of W49B displays similar features as G42.43-0.26 and G43.19-0.53 at the $\sim$10 km s${}^{-1}$ and no absorption at $\sim$12 km s${}^{-1}$ (also see the {\HI} channel map in Figure \ref{fig:3}), so reasonably its absorption at 10 km s$^{-1}$ likely originates from the local {\HI} gas. The lack of absorption from Perseus spiral arm hints its distance is likely closer than W49A. To confirm this, we calculate the optical depth assuming W49B is located at the same distance as or behind W49A, i.e. 11.4 kpc. The brightness temperature of {\HI} emission at velocity of $\sim$12 km s${}^{-1}$ in the direction of W49B is ${T}_{b} =$ 70 - 80 K. Assuming the background continuum emission is very low (${T_{bg}} = 0$ K) and the spin temperature of {\HI} cloud (${T_s}$) ranges from 90 to 300 K. The Radiative transfer equation gives the {\HI} optical depth ${\tau}_{HI} = - \ln (1 - \frac{{{T_b}}}{{{T_s}}})$, so we derive a ${\tau}_{HI}$ of 0.3 - 2.2 which is above the detection sensitivity of 0.1. Since no absorption at $\sim$12 km s$^{-1}$ is detected, W49B should be closer than W49A. Further analysis will be presented in the discussion section.\\

\subsection{{\it Spitzer} IRS low resolution spectrum}

\begin{figure}[!htpb]

\centerline{\includegraphics[width=0.45\textwidth, angle=0]{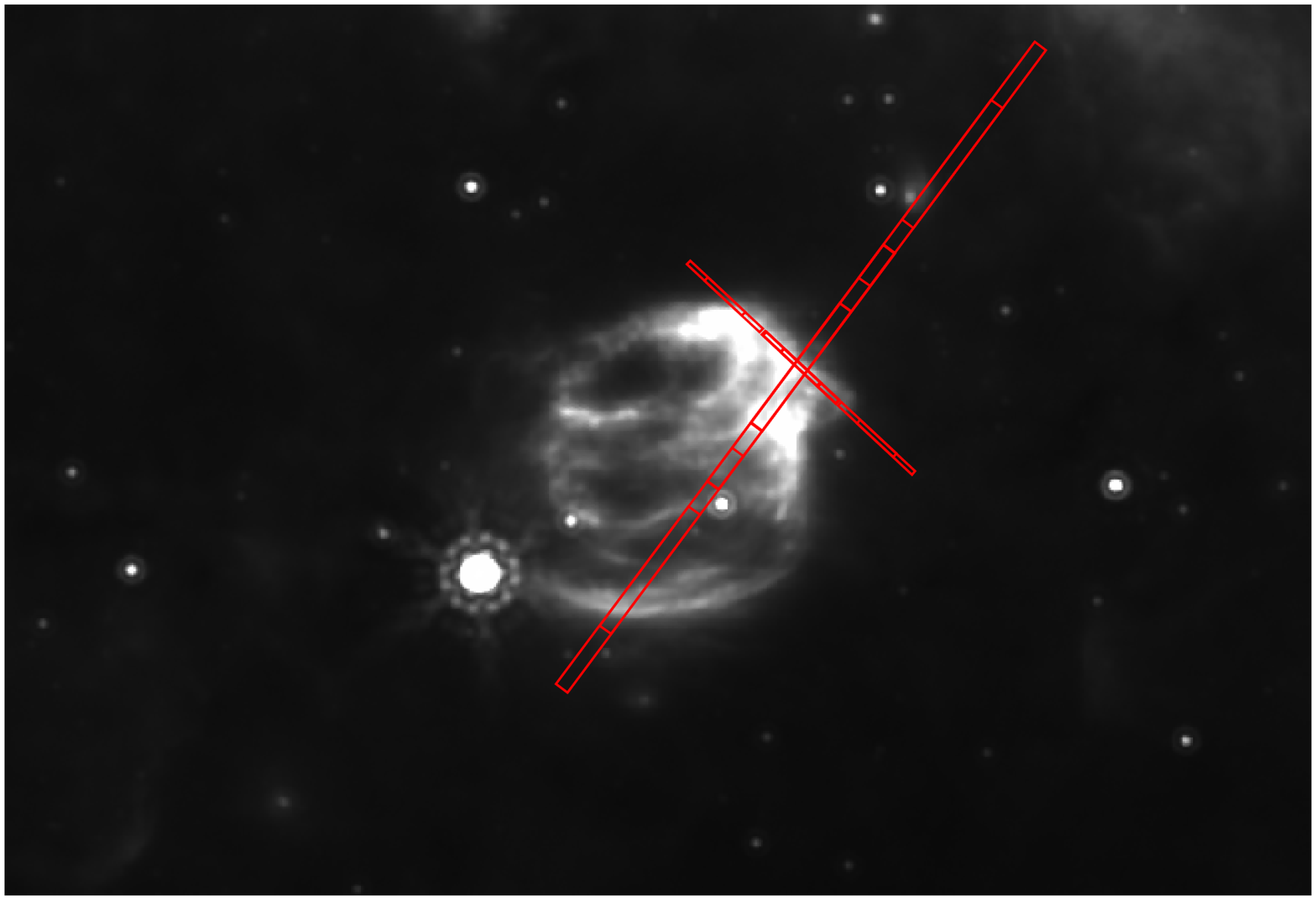}}
\caption{{\it Spitzer} 24 $\mu$m image of W49B in Galactic coordinate. Narrow rectangles represent the SL slits and broad rectangles are the LL slits.}
\label{fig:4}
\end{figure}

\begin{figure}[!htpb]
\centerline{\includegraphics[width=0.45\textwidth, angle=0]{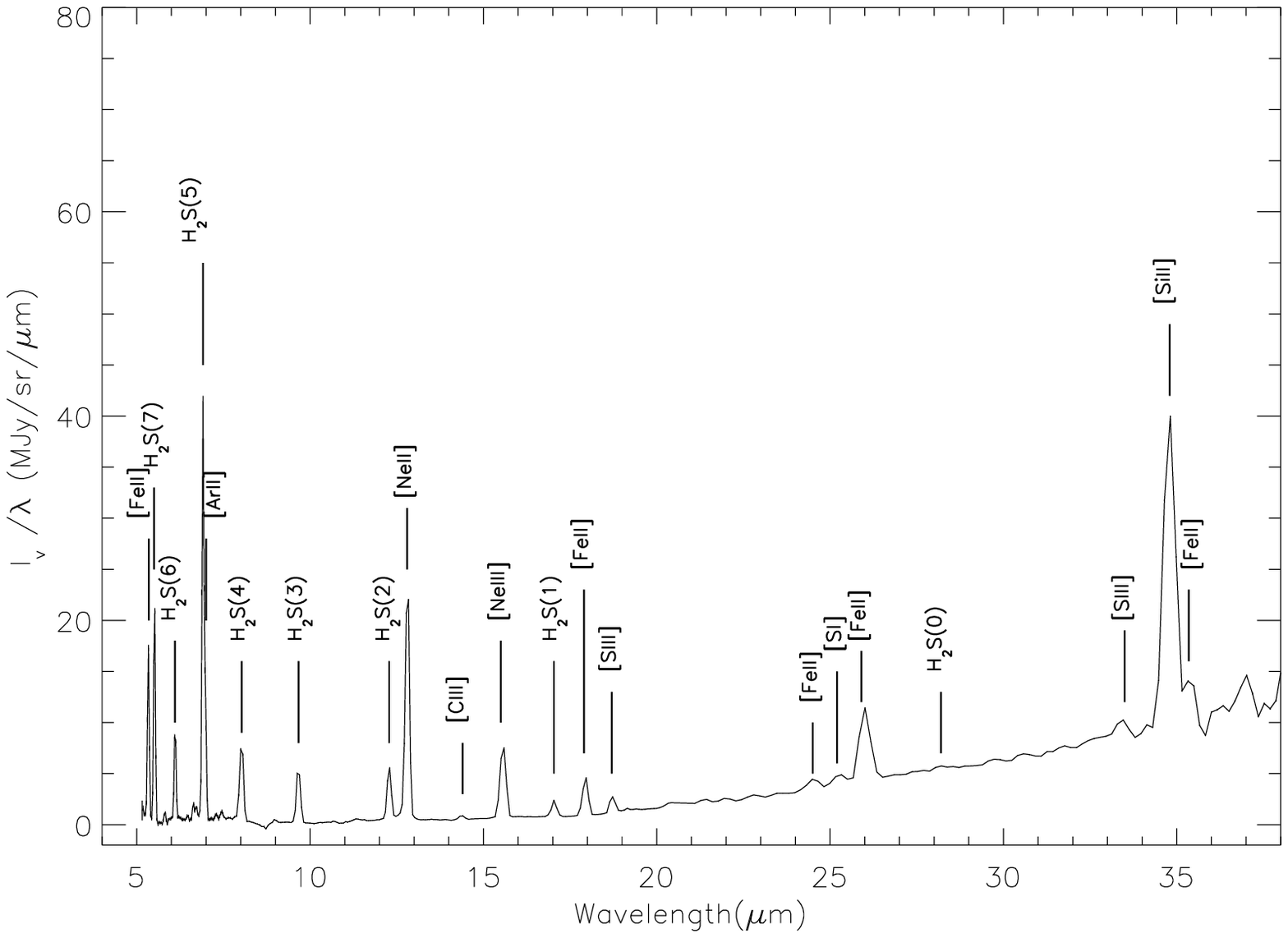}}
\caption{IRS spectrum of W49B.}
\label{fig:5}
\end{figure}

\noindent Figure \ref{fig:4} shows the IRS map of W49B. Short boxes represent the SL slits and the long boxes are for the LL slits. To correct the extinction, we use the dust model of \citet*{dra03} with ${R}_{V} = 3.1$. The selected hydrogen column density of 4.8 $\times$ 10${}^{22}$ $\textrm{cm}^{-2}$ is the average of several X-ray observations' results (e.g. \citealt{hwa00}, \citealt{mic06}, \citealt{keo07}). The final spectrum is presented in Figure \ref{fig:5}.\\

We calculate the flux of each spectral line through a gauss fitting before and after extinction correction. The results are listed in Table \ref{tab:1}. As displayed in Figure \ref{fig:5} and Table \ref{tab:1}, pure rotational transition lines of H$_2$ (0,0) S(0)-S(7) are detected. Since these bright lines usually appear in the shocked region, this supports that W49B is interacting with molecular clouds at its southwest boundary. Beside the molecular lines, forbidden transition lines of atom and ions are also detected, i.e. lines from S$^0$, S$^{2+}$, Ar$^{+}$, Cl$^{+}$, Fe$^{+}$, Ne$^{+}$, Ne$^{2+}$ and Si$^{+}$. Detecting spectral lines from different gas phase suggests W49B has a complex environment.\\

\begin{table}[t]
\scriptsize
\centering
\caption{Surface Brightness of Emission Lines of W49B}
\begin{tabular}[t]{cccc}

\hline
\hline
Transition   & $\lambda$ & Observed Brightness & De-reddened Brightness   \\
             &   $\mu$m  & erg cm${}^{-2}$ s${}^{-1}$ sr${}^{-1}$ & erg cm${}^{-2}$ s${}^{-1}$ sr${}^{-1}$ \\
\hline
H$_2$ S(7) & 5.51 & 7.56(0.92)E-4 & 1.22(0.14)E-3 \\
H$_2$ S(6) & 6.11 & 2.99(0.42)E-4 & 4.74(0.63)E-4 \\
H$_2$ S(5) & 6.91 & 1.14(0.13)E-3 & 1.64(0.19)E-3 \\
H$_2$ S(4) & 8.03 & 3.85(0.49)E-4 & 7.71(0.93)E-3 \\
H$_2$ S(3) & 9.67 & 2.35(0.28)E-4 & 1.59(0.18)E-3 \\
H$_2$ S(2) & 12.28 & 1.82(0.22)E-3 & 3.83(0.43)E-4 \\
H$_2$ S(1) & 17.04 & 4.88(0.64)E-5 & 9.35(1.17)E-5 \\
H$_2$ S(0) & 28.22 & 1.05(0.45)E-5 & 1.55(0.65)E-5 \\
$[$Fe\, II] & 5.34 & 5.89(0.78)E-4 & 9.76(1.24)E-4 \\
$[$Ar\, II] & 6.98 & 3.37(0.56)E-4 & 4.85(0.77)E-4 \\
$[$Ne\, II] & 12.8 & 7.95(0.85)E-4 & 1.48(0.16)E-3 \\
$[$Cl\, II] & 14.37 & 1.63(0.34)E-5 & 2.48(0.49)E-5 \\
$[$Ne\, III] & 15.5 & 2.93(0.31)E-4 & 4.94(0.52)E-4 \\
$[$Fe\, II] & 17.9 & 1.22(0.14)E-4 & 2.45(0.27)E-4 \\
$[$S\, III] & 18.7 & 4.97(0.65)E-5 & 1.00(0.12)E-4 \\
$[$Fe\, II] & 24.5 & 4.10(0.54)E-5 & 5.57(0.73)E-5 \\
$[$S\, I] & 25.2 & 4.12(0.55)E-5 & 6.50(0.84)E-5 \\
$[$Fe\, II] & 25.9 & 3.16(0.34)E-4 & 4.78(0.50)E-4 \\
$[$S\, III] & 33.5 & 4.80(0.71)E-5 & 6.96(0.98)E-5 \\
$[$Si\, II] & 34.8 & 1.13(0.11)E-3 & 1.51(0.16)E-3 \\
$[$Fe\, II] & 35.35 & 2.77(0.84)E-4 & 3.75(1.14)E-4 \\
\hline

\end{tabular}
\label{tab:1}
\end{table}

\subsection{Dust revealed by multi-infrared image}

\begin{figure}[!htpb]

\centering{\includegraphics[width=0.45\textwidth, angle=0]{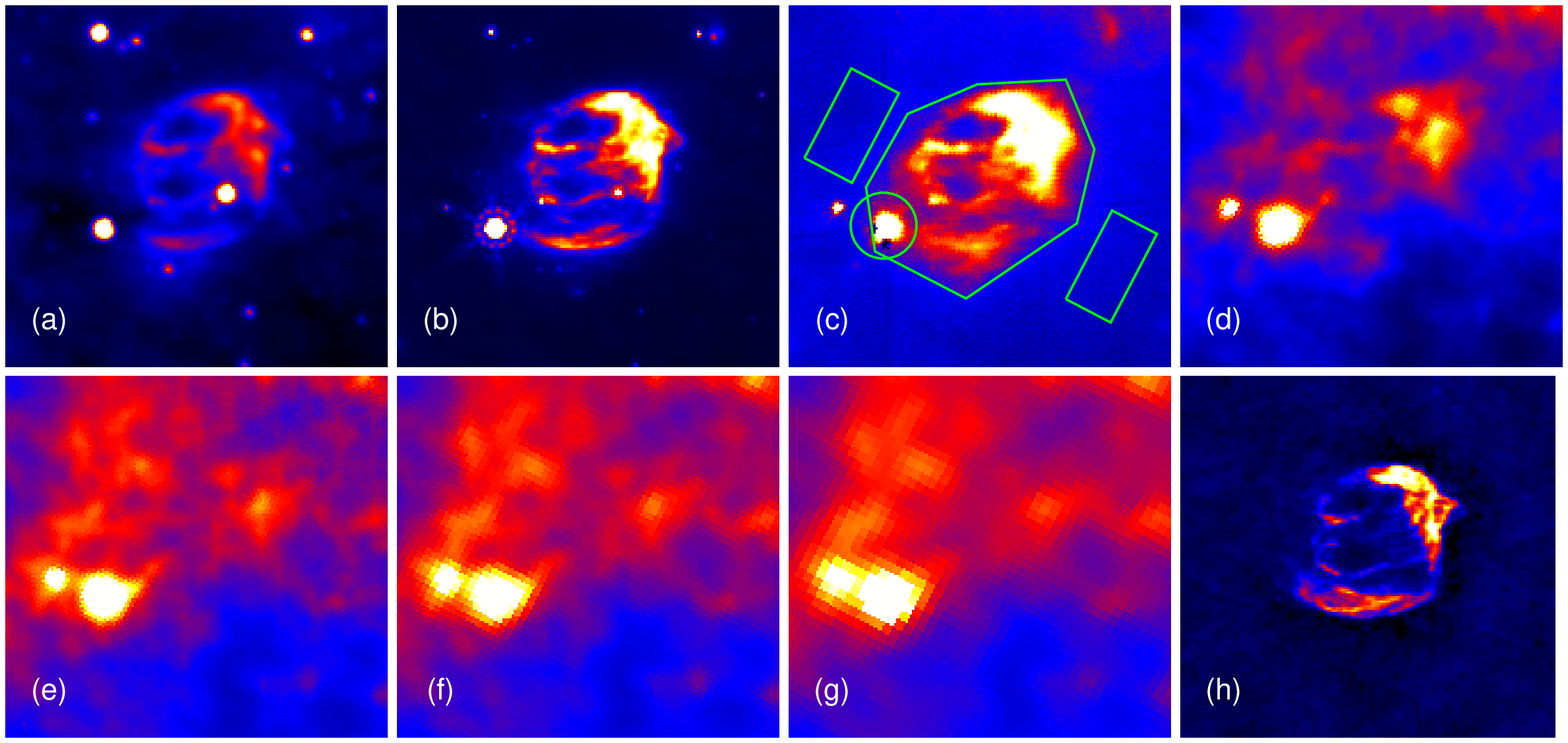}}
\caption{Infrared and radio images of W49B in Galactic coordinate. (a): 12 $\mu$m from {\it WISE}; (b): 24 $\mu$m from {\it Spitzer}; (c)-(g): 70 $\mu$m, 160 $\mu$m, 250 $\mu$m, 350 $\mu$m, 500 $\mu$m from {\it Herschel}; (h): 20 cm from MAGPIS. The region within the polygon but outside the circle is used to extract the total flux. The two rectangles are used to subtract the local background.}

\label{fig:6}
\end{figure}

Figure \ref{fig:6} presents the middle and far infrared images (from the {\it WISE}, {\it Spitzer} and {\it Herschel}) and the 20 cm continuum image from MAGPIS. W49B's morphology in the infrared images with wavelength less or equal to 70 $\mu$m is similar with that in the radio image which indicates that the infrared emissions are correlated with W49B. At 160 $\mu$m, unrelated foreground and background cold dust emissions begin to dominate the image, but weak features correlating with the radio emission can still be seen. For longer wavelengths, infrared emissions related with W49B entirely disappear in the images.\\

\begin{figure}[!htpb]

\centerline{\includegraphics[width=0.45\textwidth, angle=0]{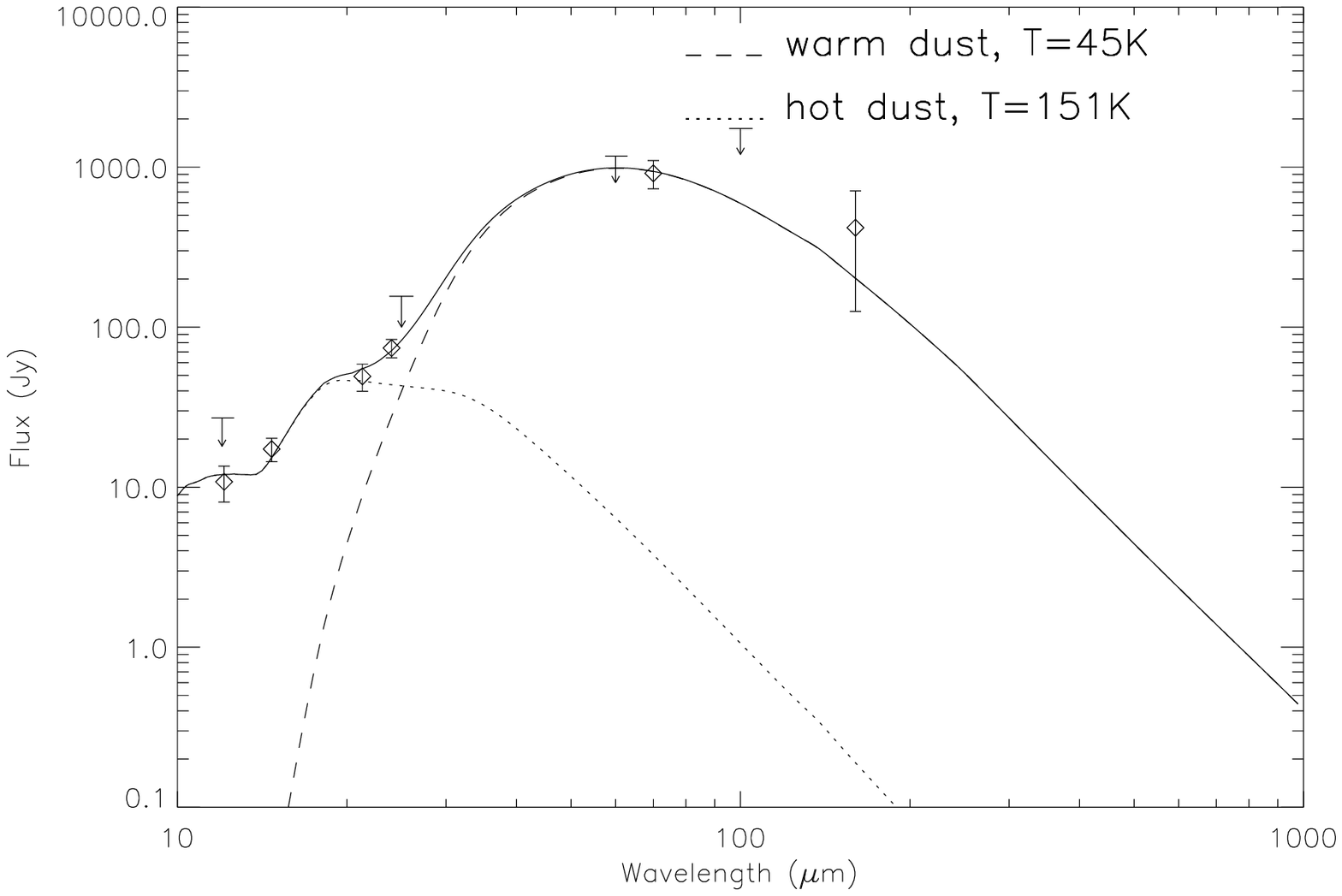}}
\caption{Mid- and far-infrared SED of W49B. The dotted line is for the hot component. The dashed line is for the warm component. The solid line is the sum of two components. The upper limits at 12, 25, 60 and 100 $\mu$m are from the Improved Reprocessing of the IRAS Survey (\citealt*{miv05}).}

\label{fig:7}
\end{figure}

We fit the spectral energy distribution (SED) to understand the property of dusts related with W49B. In the Figure \ref{fig:6}c, we select the region within the polygon excluding the circle to extract the emission fluxes from 12 $\mu$m to 160 $\mu$m. Two rectangles are used to subtract the local background. Because W49B shows strong spectral lines (see Figure \ref{fig:5}) which contribute important emissions to the middle infrared images, we correct the line contributions by convolving the IRS spectrum with the response curves of {\it Spitzer} and {\it MSX}. Because the synchrotron flux of W49B at 6 cm is about 38 Jy and the spectral index is about -0.48 (\citealt*{mof94}), the contribution of electron synchrotron emission to the infrared flux is negligible (The contributed fluxes from 160 $\mu$m to 12.1 $\mu$m are 2.21, 1.49, 0.89, 0.84, 0.70 and 0.64 Jy). The fluxes at each wavelength are listed in Table \ref{tab:2}. We fit a two-component modified blackbody to the dust SED. The function is:\\
\begin{equation}
{S_v} = \frac{{{M_h}}}{{{D^2}}} \times {\kappa _v} \times B(v,{T_h}) + \frac{{{M_w}}}{{{D^2}}} \times {\kappa _v} \times B(v,{T_w}),
\end{equation}
where ${M_h}$ and ${M_w}$ represent the masses of the hot and warm dust, $B(v,T)$ is the Plank function and ${{\kappa}_{v}}$ is the dust absorption cross section per unit mass which is from \citet*{dra03} with ${R_v} = 3.1$. Figure \ref{fig:7} displays the result. The hot dust has a temperature of 151 $\pm$ 20 K with mass of 7.5 $\pm$ 6.6 $\times$ ${10}^{-4}$ ${\Msol}$ and the warm dust has a temperature of 45 $\pm$ 4 K with mass of 6.4 $\pm$ 3.2 ${\Msol}$. The dust mass of W49B is calculated using a distance of 10 kpc (see the discussion section for distance).\\

\begin{table}[t]
\scriptsize
\centering
\caption{Fluxes of each wavelength used for SED fitting}
\begin{tabular}[t]{ccccccc}
\hline
\hline
Wavelength ($\mu$m) & 12.1  & 14.7  & 21.3  & 24.0  & 70.0   & 160.0  \\
\hline
Flux (Jy)           & 10.80 & 17.30 & 49.29 & 74.06 & 915.64 & 418.25 \\
Error (Jy)           & 2.73  & 2.89  & 9.50  & 9.80  & 183.13 & 292.78 \\
\hline
\end{tabular}
\label{tab:2}
\end{table}

\section{Discussions}

\subsection{The distance of W49B}

\begin{figure}[!htpb]

\centerline{\includegraphics[width=0.45\textwidth, angle=0]{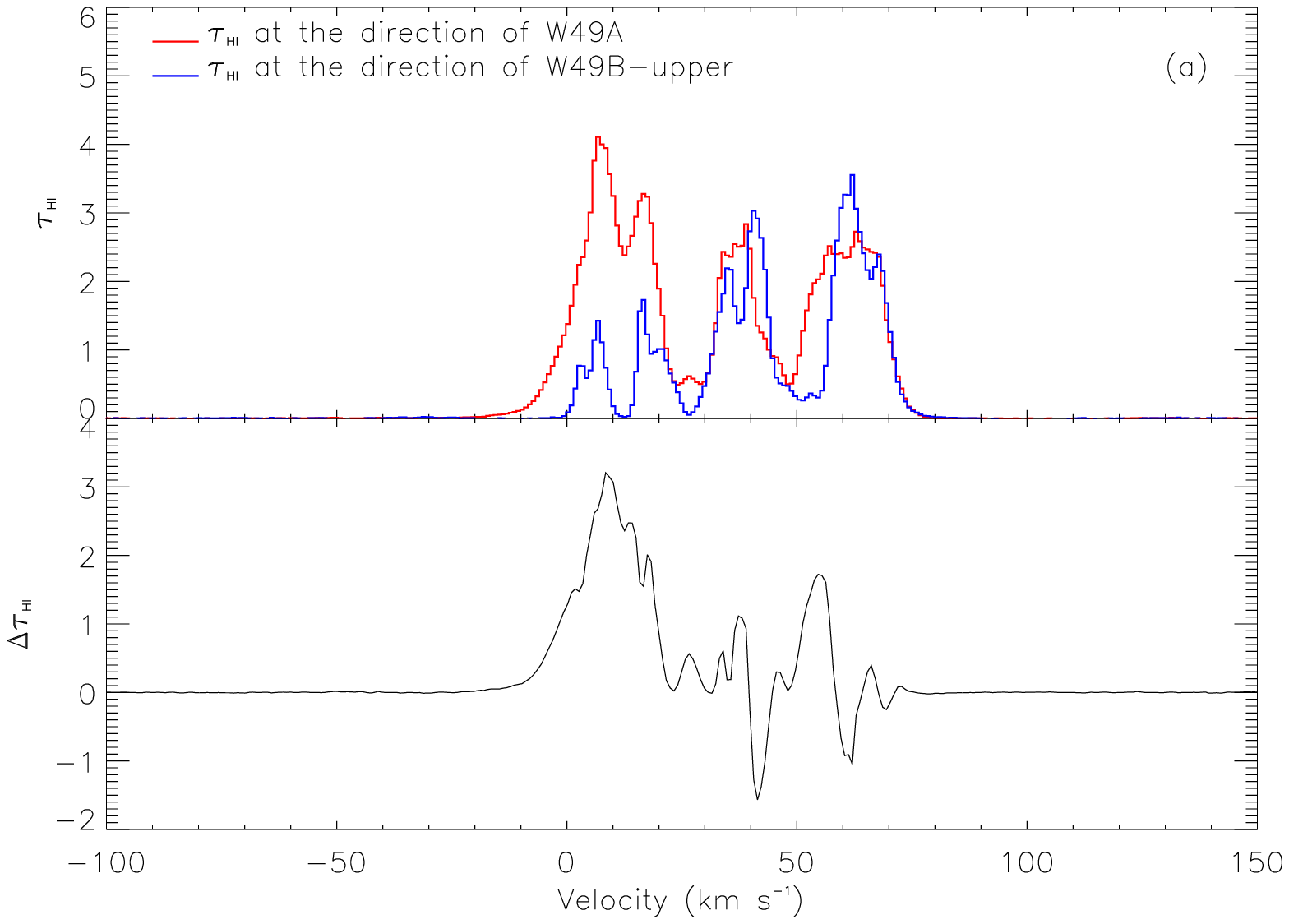}}
\centerline{\includegraphics[width=0.45\textwidth, angle=0]{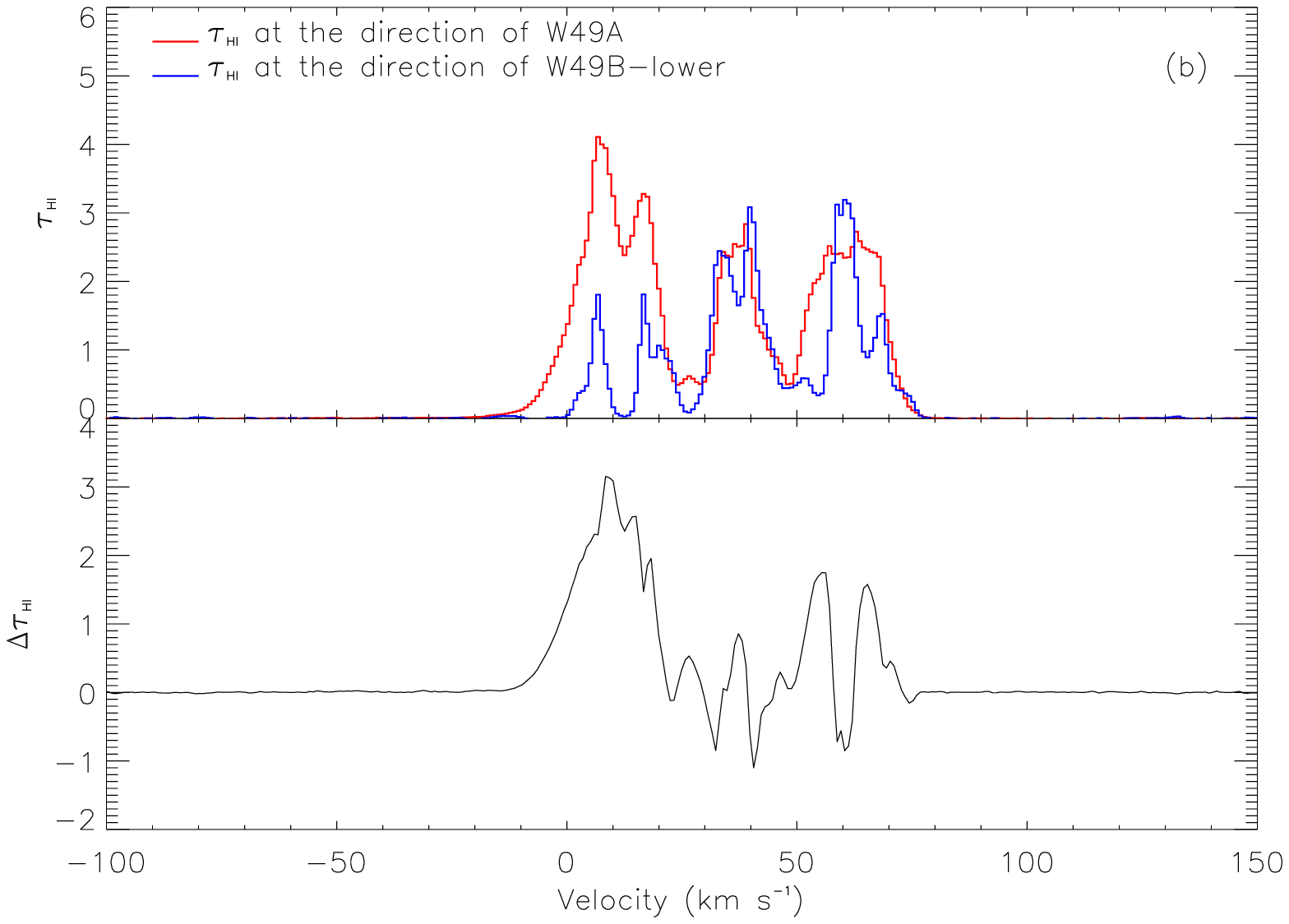}}
\centerline{\includegraphics[width=0.45\textwidth, angle=0]{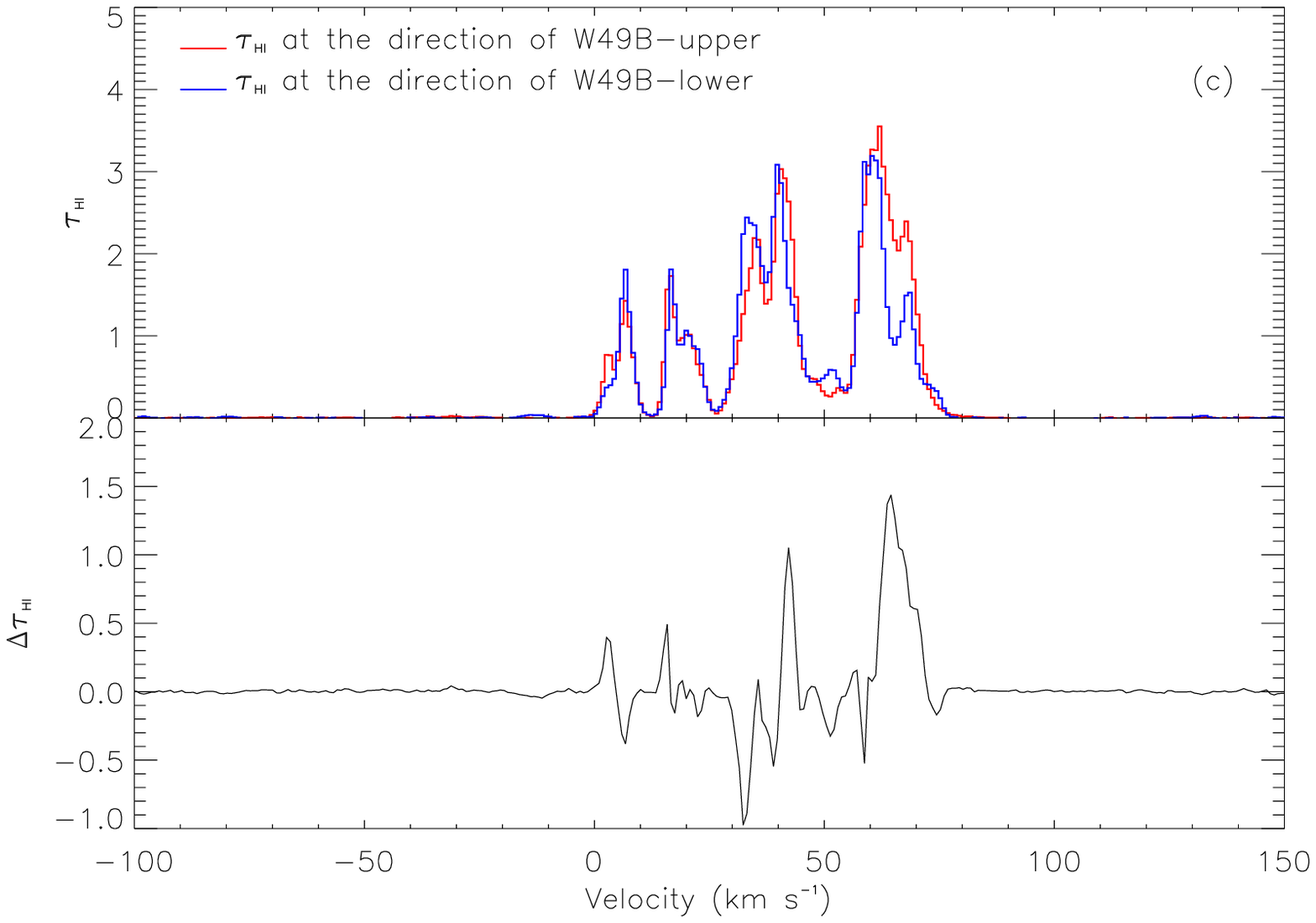}}
\caption{{\HI} optical depth towards W49B and W49A.}

\label{fig:8}
\end{figure}

\noindent Previous {\HI} absorption studies suggested W49B at a distance of $\sim$8.0 kpc (\citealt*{mof94}) or $\sim$11.4 kpc (\citealt*{bro01}). The distance of 8.0 kpc is based on two reasons. The first one is that {\HI} absorptions are seen clearly up to velocity of $\sim$70 km s${}^{-1}$ which argues W49B lies behind the tangent point at $\sim$6.2 kpc (assuming Galactic center distance ${R_\odot} = 8.5$ kpc and rotation velocity ${V_\odot} = 220$ km s${}^{-1}$). The second one is that, compared with W49A, W49B is absent {\HI} absorptions at velocities from 7 to 14 km s${}^{-1}$ and 50 to 55 km s${}^{-1}$. The lack of absorption at $\sim$55 km s$^{-1}$ favors W49B is at the distance of $\sim$8.0 kpc. We contrast the {\HI} optical depth (${\tau}_{HI}$) of W49B with W49A in Figure \ref{fig:8}. The ${\tau}_{HI}$ differences at velocities from 7 to 14 km s${}^{-1}$ and 50 to 55 km s${}^{-1}$ are confirmed. Beside them, there are other two ${\tau}_{HI}$ differences at velocities of $\sim$40 km s${}^{-1}$ (Figure \ref{fig:8}a) and $\sim$65 km s${}^{-1}$ (Figure \ref{fig:8}b) which have similar intensity as the $\Delta{\tau}_{HI}$ at $\sim$55 km s${}^{-1}$. The {\HI} absorption of different regions of W49B has obvious difference at $\sim$65 km s${}^{-1}$ (Figure \ref{fig:8}c). Moreover, in the ${N_{HI}}/{T_s}$ image toward both W49A and W49B, \citet*{bro01} found there is obvious change on size scales of about 1${'}$. Therefore, the reason supporting the distance of $\sim$8.0 kpc for W49B is not sufficient.\\

\begin{figure}[!htpb]

\centerline{\includegraphics[width=0.45\textwidth, angle=0]{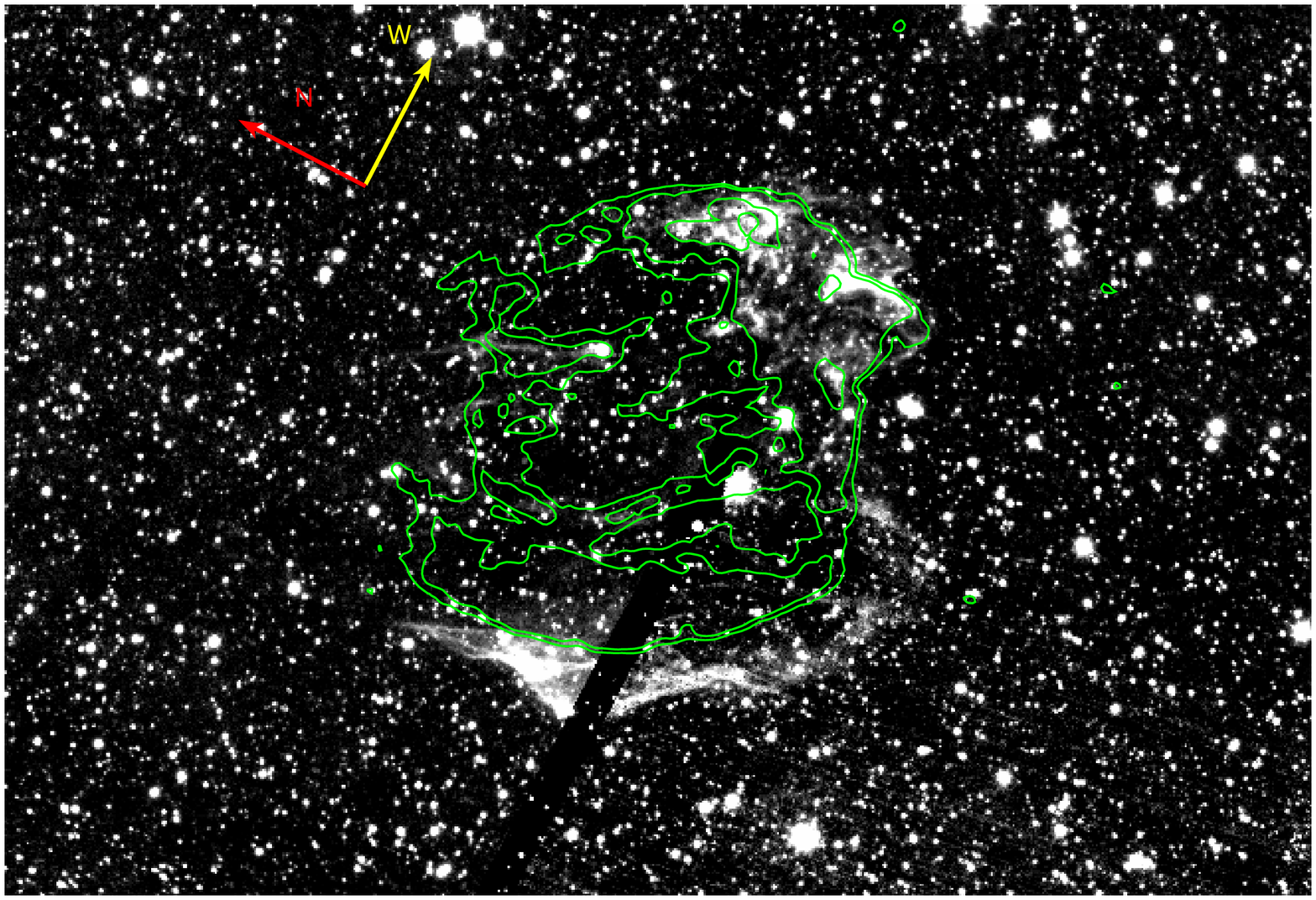}}
\caption{2.12 $\mu$m image of W49B in Galactic coordinate. The contours is from 20 cm image of MAGPIS with levels of 3.5, 10, 50 and 100 mJy/beam. The red arrow is for the direction of north and the yellow arrow represents west.}

\label{fig:9}
\end{figure}

\begin{figure*}[!htpb]
\centerline{\includegraphics[width=0.33\textwidth, angle=0]{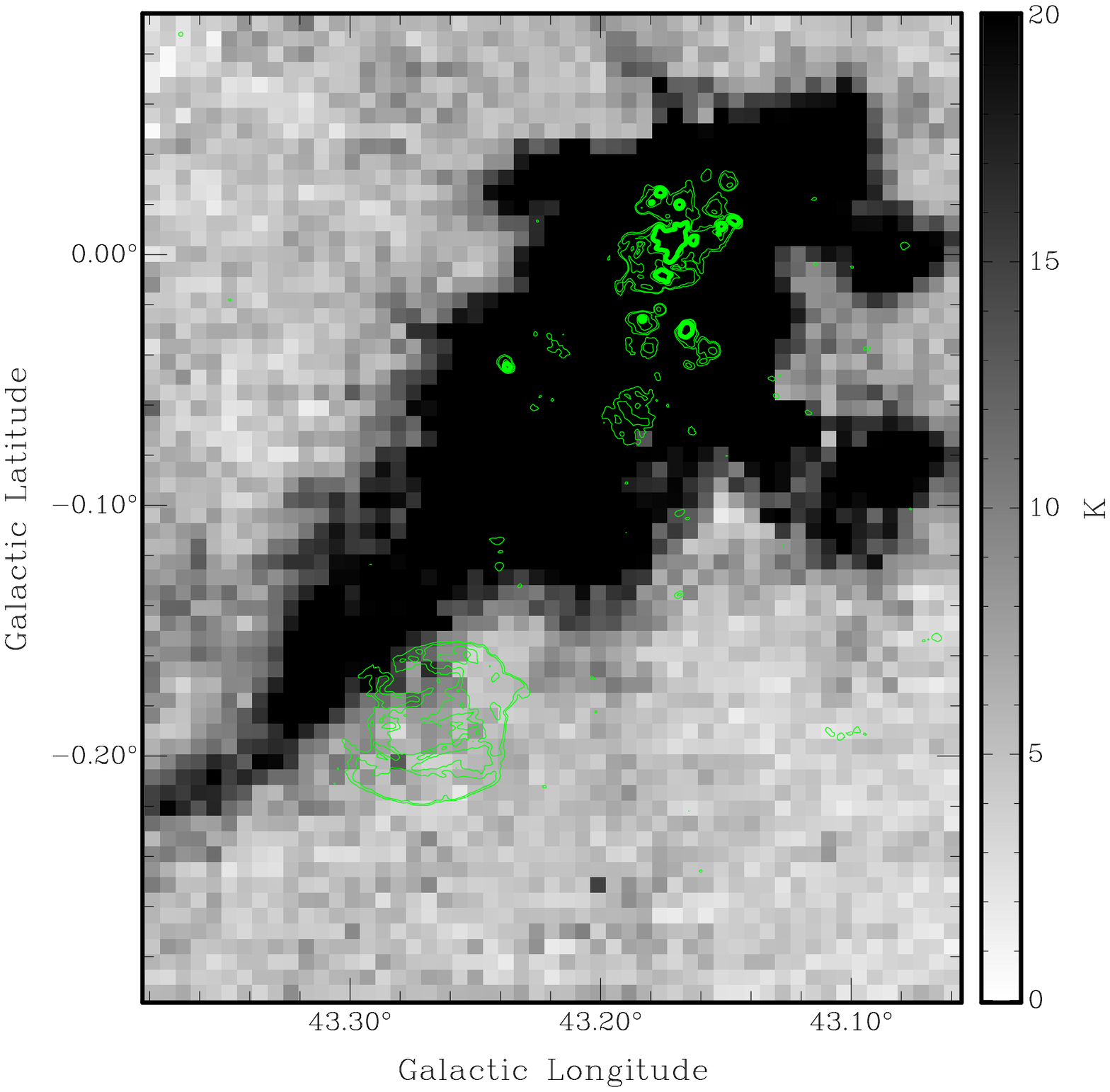}\includegraphics[width=0.33\textwidth, angle=0]{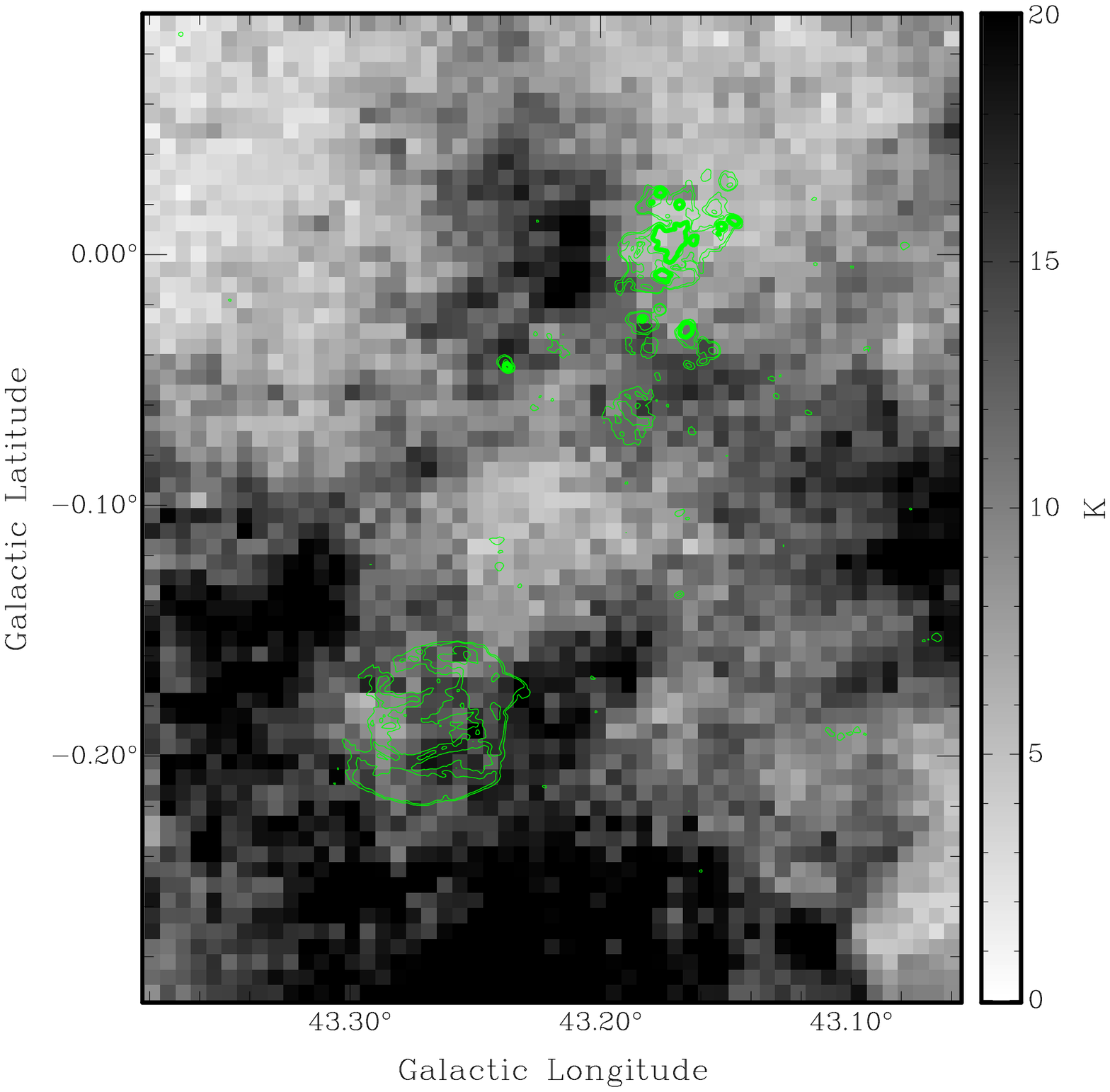}\includegraphics[width=0.33\textwidth, angle=0]{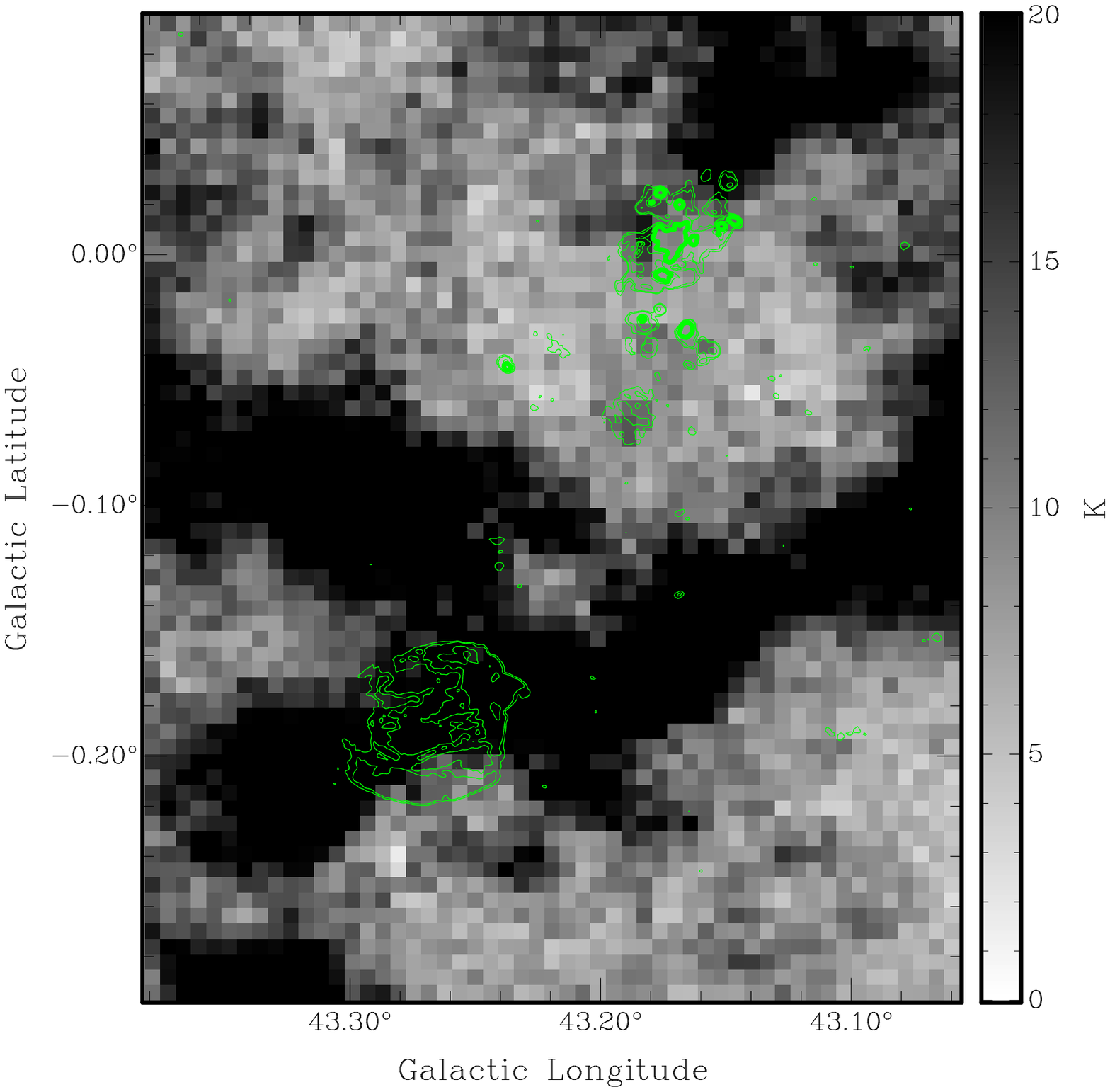}}
\centerline{\includegraphics[width=0.33\textwidth, angle=0]{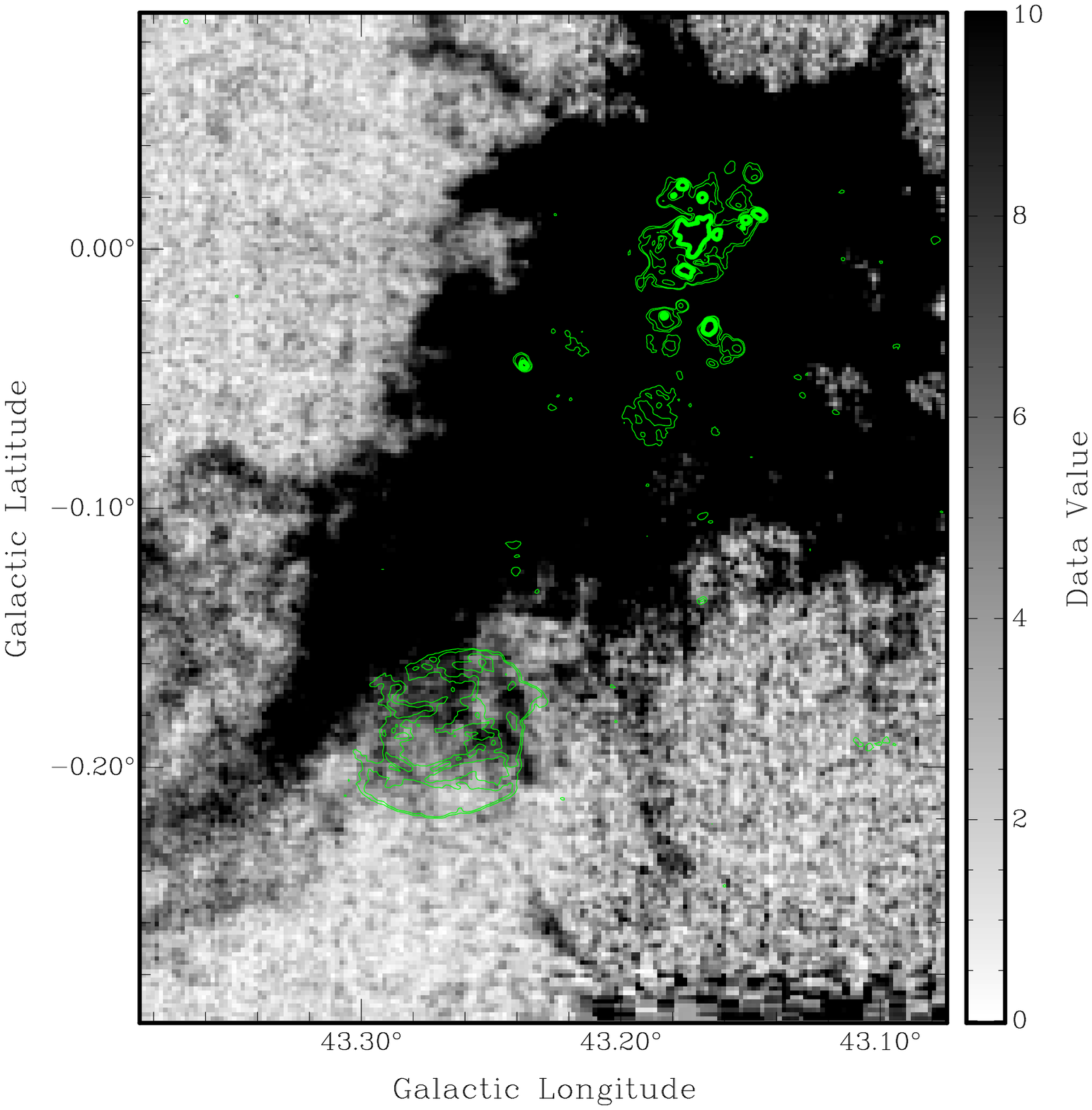}\includegraphics[width=0.33\textwidth, angle=0]{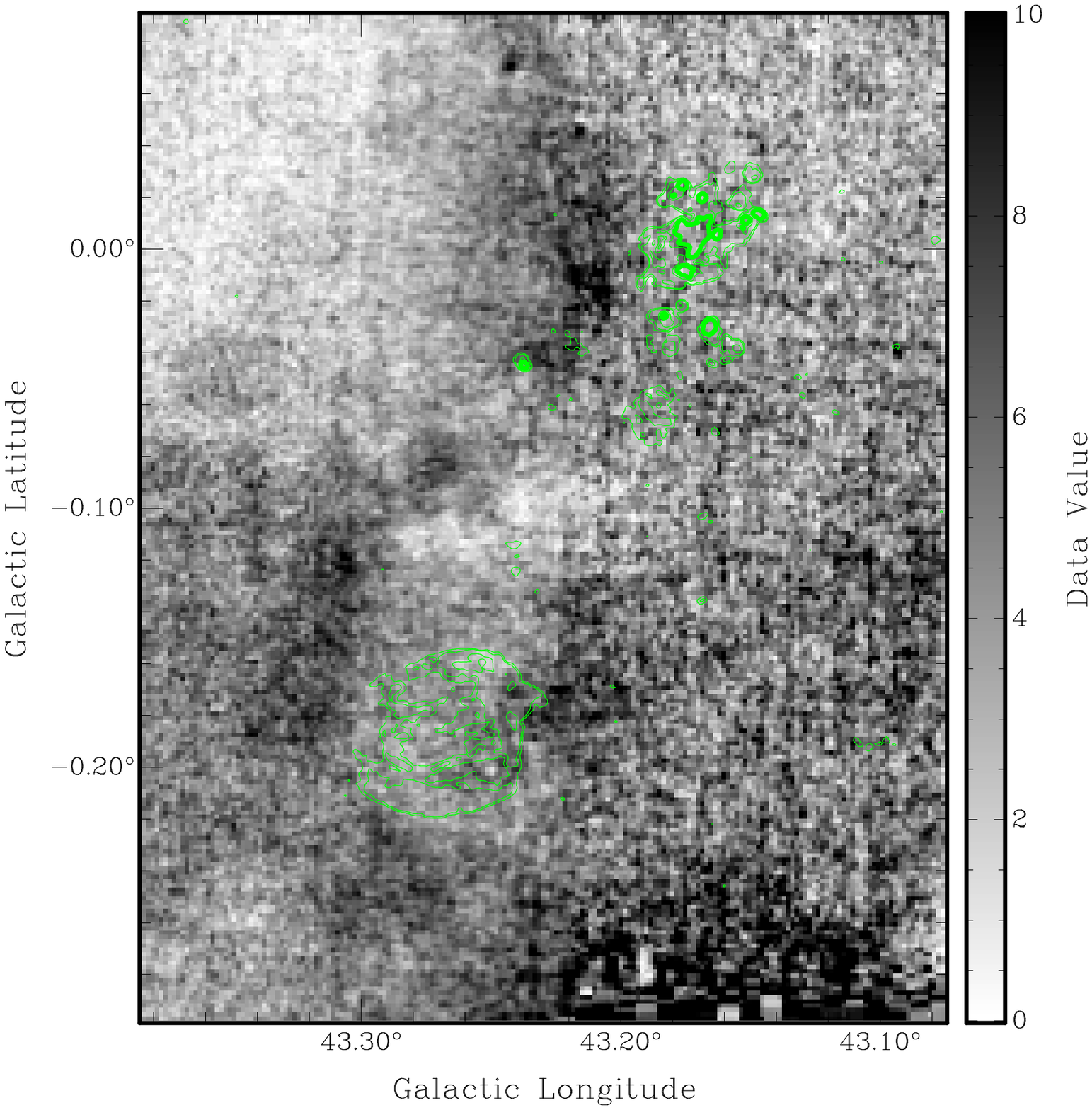}\includegraphics[width=0.33\textwidth, angle=0]{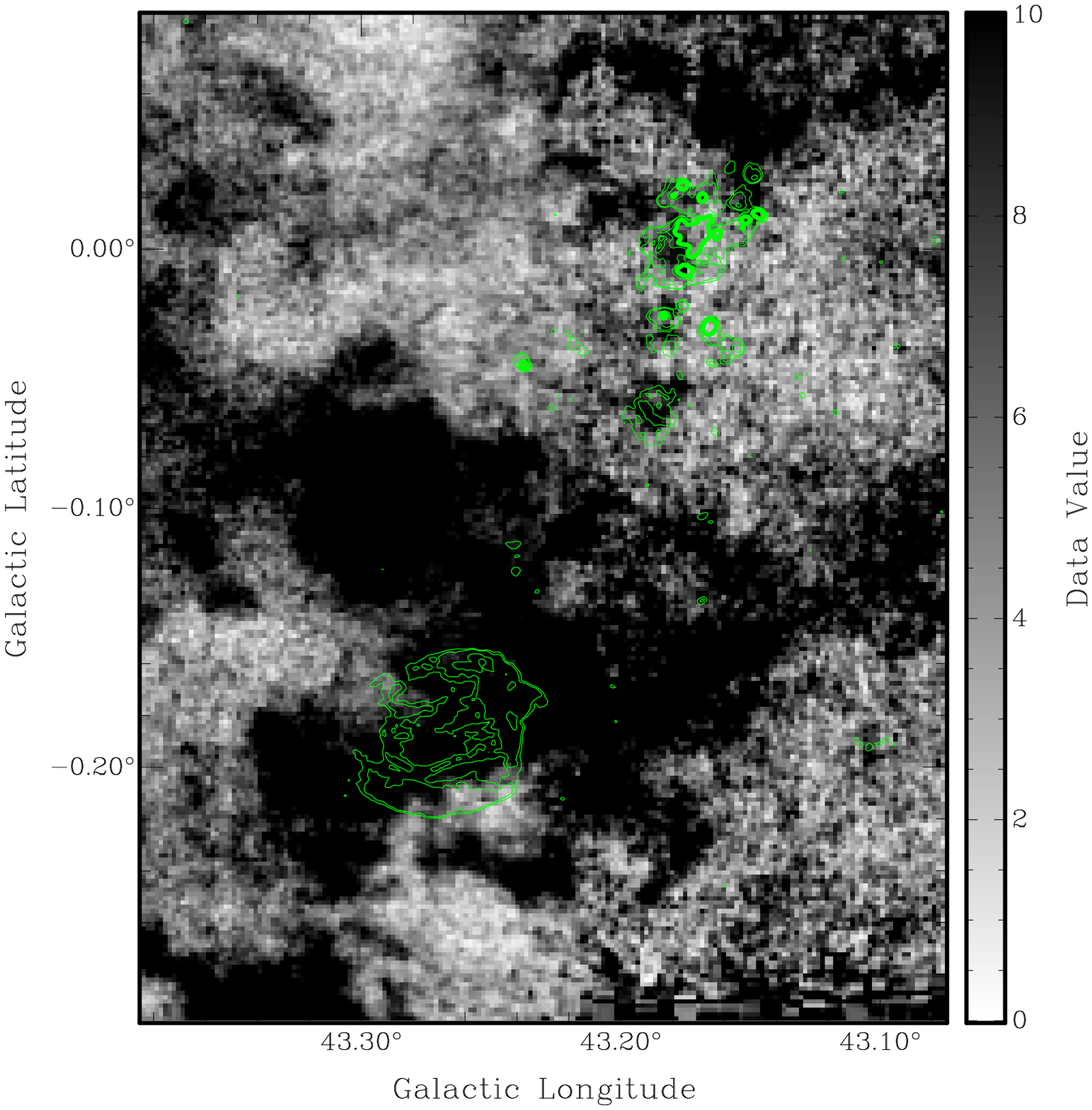}}
\caption{Upper: ${}^{13}\textrm{CO J=1-0}$ intensity maps integrated from 1-15 km s${}^{-1}$, 38-47 km s${}^{-1}$ and 57-67 km s${}^{-1}$. Lower: $\textrm{CO J=3-2}$ intensity maps integrated from the same velocity range as ${}^{13}\textrm{CO J=1-0}$ map. The contours in each panel have the same levels as the contour in Figure \ref{fig:9}.}
\label{fig:10}
\end{figure*}

The evidence for distance of 11.4 kpc is weak. Figure 9e of \citet*{bro01} shows the ${N_{HI}}/{T_s}$ image of W49B integrated velocities from -4.8 to 12.6 km s${}^{-1}$. They found distinct {\HI} enhancement concentrated toward the south boundary of the SNR. Because the enhancement is coincident with the sharp edge in the 20 cm continuum image, they explained it as the result of interaction between W49B and its ambient medium. If it was true, W49B would locate at the far side distance of $\sim$5 km s$^{-1}$, i.e. $\sim$12 kpc which is nearly the same distance as W49A. For the {\HI} absorption differences at velocities from 7 to 14 km s${}^{-1}$ between W49A and W49B, they suggested it might be caused by the differences of {\HI} kinematics, distribution and temperature in the direction of W49B and W49A. But as illustrated in section 3.1, even with large scale variations of spin temperature, {\HI} absorption at velocity of $\sim$12 km s${}^{-1}$ should be visible if W49B is at 11.4 kpc. There are other evidences supporting that W49B is closer than W49A. Figure \ref{fig:9} shows the 2.12 $\mu$m image which traces the shocked H$_2$. The left panels of Figure \ref{fig:10} present the intensity maps of ${}^{13}\textrm{CO} (J=1-0)$ and $\textrm{CO} (J=3-2)$ integrated from velocity of  1 - 15 km s${}^{-1}$. The shocked H$_2$ can be seen clearly at the east, south and west boundaries of W49B, indicating molecular clouds surround W49B except the northern part. This picture is also consistent with the radio continuum image (Figure 4 of \citealt*{mof94}, Figure 2 of \citealt{lac01}) which shows sharp boundary at the east, south and west and diffuse emission at the north. Diffuse emission means long electron free path length so low medium density. On the contrary, CO clouds at $\sim$10 km s${}^{-1}$ only show strong emission at north. W49B is likely not associated with the 10 km s${}^{-1}$ CO clouds so is not at the same distance of W49A.\\

Since the [Fe II] 1.64 $\mu$m, the H$_2$ 2.12 $\mu$m morphology and the estimated mass of the progenitor prefer W49B is likely in a bubble surrounded by molecular clouds, we check the distribution of CO emission in the direction of W49B to further constrain its distance. Only the $\sim$40 km s${}^{-1}$ CO cloud has a bubble-like feature (also see Figure 1 of \citealt{che14}). Because {\HI} self-absorption is detected at velocity of 40 km s${}^{-1}$ (\citealt{sim01}), previous study suggested the CO clouds at the near side distance of 40 km s$^{-1}$. However, we find multi-velocity components in the 40 km s$^{-1}$ CO cloud. We suggest that the 40 km s$^{-1}$ CO cloud is composed of two components: the near side cloud and far side cloud, which also causes additional {\HI} self-absorption. W49B is likely surrounded by the far side CO cloud so has a distance of $\sim$10 kpc which is similar to \citet{che14}'s suggestion (9.3kpc).\\

\subsection{The property of the environment of W49B}

\subsubsection{Molecular phase}

\begin{figure}[!htpb]

\centerline{\includegraphics[width=0.45\textwidth, angle=0]{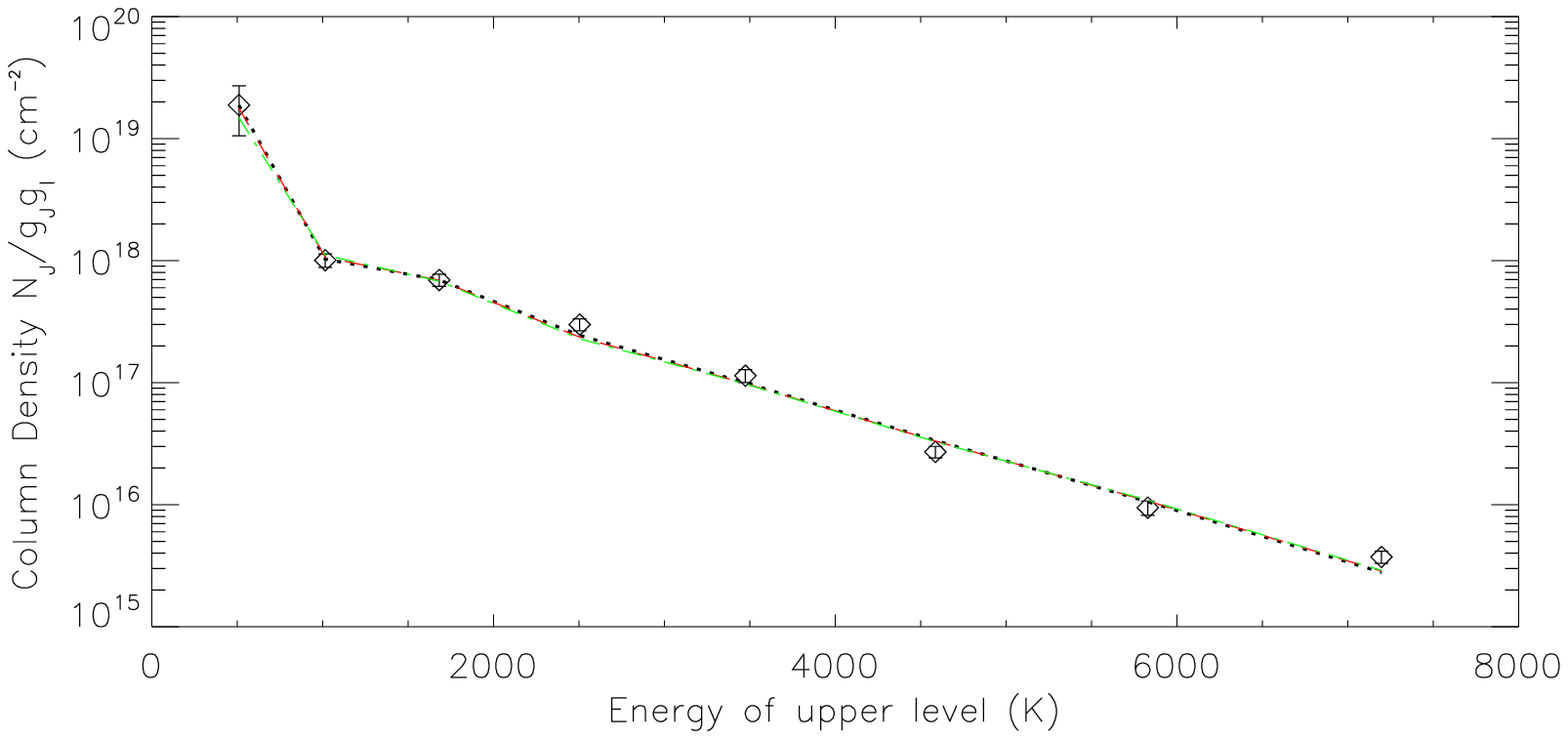}}
\caption{Boltzmann diagram of the shocked H$_2$. The black dotted line represents ${OPR}_{w}=0.01$. The red dashed line is for ${OPR}_{w}=0.05$ and the green dash-dotted line is for ${OPR}_{w}=0.1$.}

\label{fig:11}
\end{figure}

\noindent Since H$_2$ (0,0) S(0)-S(7) lines are usually optically thin, the lines' intensity can be used directly to derive the column density of the upper state of the transitions. Therefore the lines are a useful tool to estimate the temperature of the shocked H$_2$ (\citealt{hew09}). Figure \ref{fig:11} is the Boltzmann diagram of the shocked H$_2$. A striking zig-zag pattern is presented in the figure. This implies that the ortho- to para-state ratio ($OPR$) of the H$_2$ is out of equilibrium. Under the local thermodynamic equilibrium, a two temperature distributions model including parameters, ${N({H_2})_{w,h}}$ (the column density of warm and hot H$_2$ components), ${T_{w,h}}$ (the temperature) and ${OPR_{w,h}}$ (the ortho- to para-state ratio), is used to fit the data. Since a free fitting will make the ${OPR_w}$ become unreasonable small (less than ${10}^{-38}$), we fix it to be 0.01, 0.05 and 0.10. This leads to a slightly bigger ${\chi}^{2}$ than the free fitting. Our result is presented in Figure \ref{fig:11} and Table \ref{tab:3}. The three fitting lines almost overlap each other completely, indicating they have similar ${\chi}^{2}$. According to the work of \citet*{tim98}, the conversion of ortho- to para-state in the shock begins at about 700 K and quickly reachs balance with the ratio of 3 with a temperature of 1300 K. The temperature of the hot H$_2$ is $\sim$1060 K with the ortho- to para-state ratio of $\sim$0.96, this might mean the hot component is going through the transition from ortho to para.\\

\begin{table}[t]
\footnotesize
\centering
\caption{Fitted Excitation Parameters to The Shocked H$_2$ lines}
\begin{tabular}[t]{ccccccc}

\hline
\hline
Parameter          &${N({H_2})}_{w}$ & ${T}_{w}$ & ${N({H_2})}_{h}$ & ${T}_{h}$ & ${OPR}_{h}$  & ${\chi}^{2}$    \\
                   &    cm${}^{-2}$     &       K      &     cm${}^{-2}$    &       K     &     &               \\
\hline
${OPR}_{w}=0.01$& 8.86E20 &  246     &1.38E20 &   1050   &     0.97  &  14.15        \\
${OPR}_{w}=0.05$& 8.41E20 &  254     &1.32E20 &   1063   &     0.96  &  14.56         \\
${OPR}_{w}=0.10$& 6.91E20 &  266     &1.27E20 &   1075   &     0.95  &  15.60         \\
\hline

\end{tabular}
\label{tab:3}
\end{table}

\subsubsection{Ionic phase}

\begin{figure*}[!htpb]

\centerline{\includegraphics[width=0.49\textwidth, angle=0]{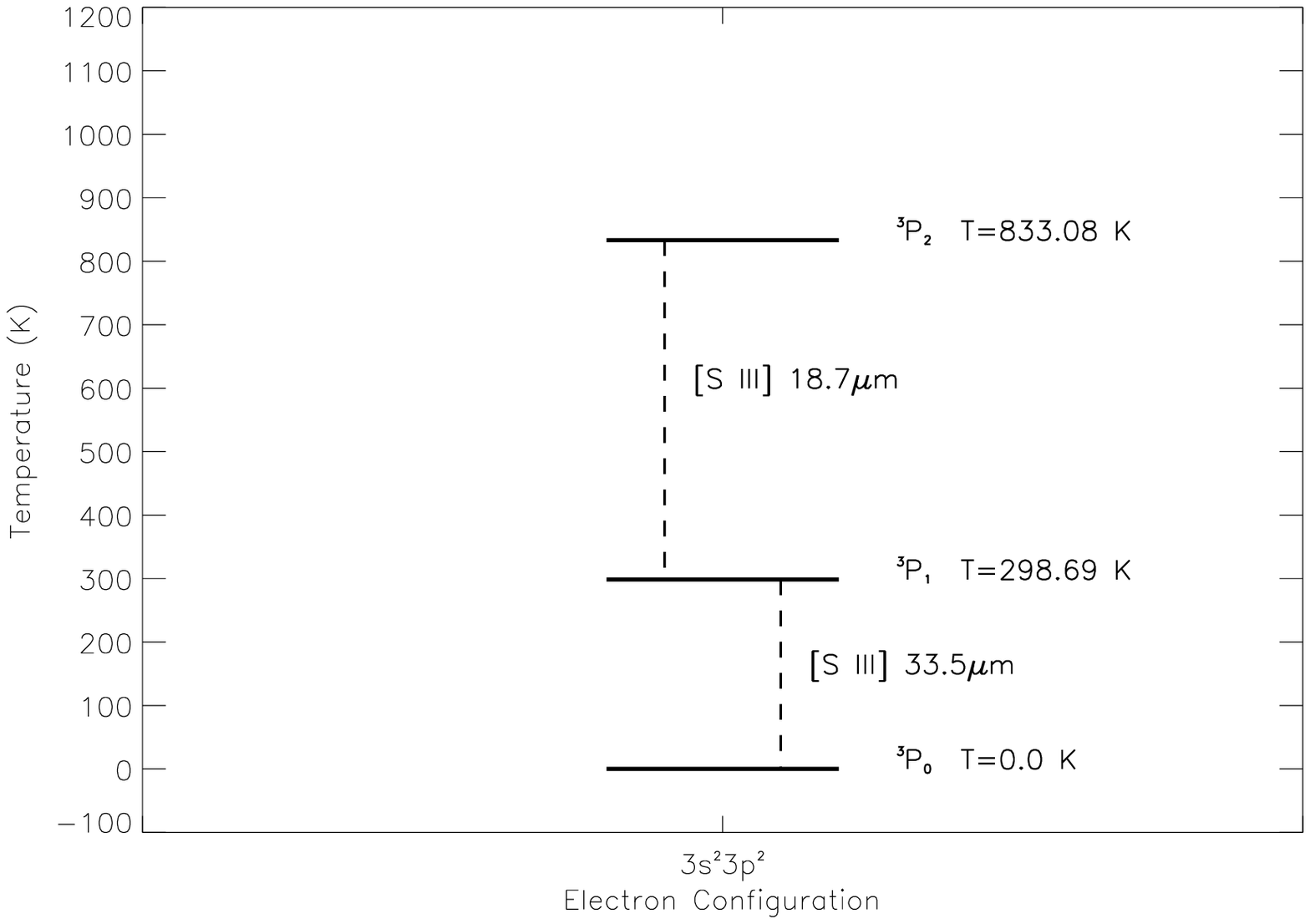}\hfill\includegraphics[width=0.49\textwidth, angle=0]{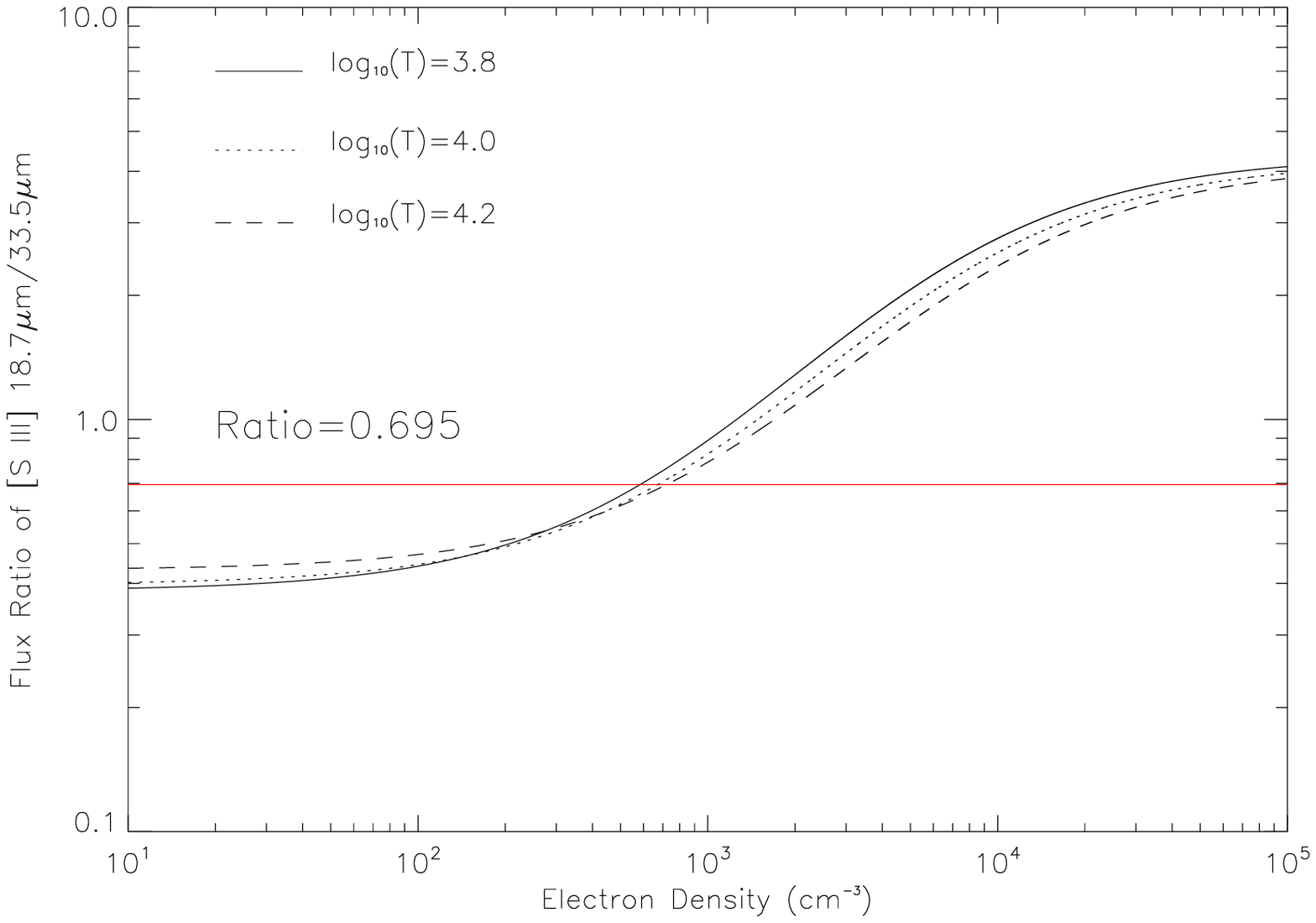}}
\caption{Left: Energy levels for $[$S III$]$ 18.7 $\mu$m and 33.5 $\mu$m line. Right: Electron density diagnosis from the flux ratio of $[$S III$]$ 18.7 $\mu$m and 33.5 $\mu$m line.}
\label{fig:12}
\end{figure*}

\noindent The abundant ionic spectral lines, especially the Fe$^{+}$ and S$^{2+}$ lines, are perfectly diagnostic tools to ionic gas. The [S III] 18.7 $\mu$m and 33.48 $\mu$m lines are produced by the fine-structure transition of ${}^{3}$P${}_{2}$ to ${}^{3}$P${}_{1}$ and ${}^{3}$P${}_{1}$ to ${}^{3}$P${}_{0}$. As showed in the left panel of Figure \ref{fig:12}, since their upper levels have similar temperature, the line intensity ratio is only sensitive to electron density. The right panel of Figure \ref{fig:12} shows the density dependence on the ratio, computed by solving the rate equations for three level systems. Transition probabilities and collision strengths are from \citet*{men82} and \citet*{gal95} respectively. The collisional deexcitation coefficient is calculated by equation(\citealt{ost06}): \\
\begin{equation}
{q_{21}} = \int\limits_0^\infty  {u{\sigma _{21}}f(u)du}  = \frac{{8.629 \times {{10}^{ - 6}}}}{{{T^{1/2}}}}\frac{{\Upsilon (1,2)}}{{{g_2}}},
\end{equation}
where u is the electron velocity, ${{\sigma _{21}}}$ represents the cross section of dedexciation, $f(u)$ is the electron velocity distribution function, ${\Upsilon (1,2)}$ is the collision strength and ${g_2}$ represents statistical weight of the upper level. The collisional excitation coefficient is derived by:\\
\begin{equation}
{q_{12}} = \frac{{{g_2}}}{{{g_1}}}e^{(h{v_{21}}/kT)}.
\end{equation}
The red line in the right panel of Figure \ref{fig:12} shows the ratio which indicates an electron density of 500-700 cm${}^{-3}$. Diagnostics of the [Fe II] lines are presented following the work of \citet{hew09} which use the excitation rate equations of \citet{rho01}. The line ratio of $17.9/5.35$ $\mu$m is sensitive primarily to electron density and the ratio of $17.9/25.9$ $\mu$m depends on both density and temperature. More details can be found in \citet{hew09}. For W49B, the former ratio is 0.25 and the latter value is 0.51. The values prefer a density of 400-600 cm${}^{-3}$ and a temperature of $\sim$10$^4$ K.\\

We notice that \citet{keo07} suggested the electron temperature and electron density surrounding W49B are range from (1000 K, 8000 cm$^{-3}$) to (700 K, 1600 cm$^{-3}$), which are different from ours. The difference may be caused by two factors. The first one is that \citet{keo07} gave the parameters only using single spectral line (i.e. [Fe II] 1.64 $\mu$m) measurement based on some assumptions by considering a wide range of reasonable excitation conditions, but still not covering all possible conditions for SNR (see table 5, 9 of \citealt{hew09} and table 3 of \citealt{neu07}), this might cause large uncertainty for the parameters comparing with ours based on a pair of spectral lines which are powerful diagnostic tools for the measurement. The second one is that we might observe parts of W49B different from theirs. As revealed by the {\it Spitzer} IRS spectrum, W49B has a very complex environment including all three phases gas (the molecular, neutral and ionic gas). In large scale, the temperature and density distribution may form a gradient descent from inner to outside due to the stellar wind before supernova explosion and the interaction between the SNR and its surrounding cloud. If the IRS observation point is closer to inner part compared with \citet{keo07}'s, the electron temperature and density should be higher and more tenuous.\\

To clarify this explanation, we need compare the spatial distribution of [S III] 18.7 $\mu$m and 33.48 $\mu$m emission, [Fe II] 5.34 $\mu$m, 17.9 $\mu$m and 25.9 $\mu$m emission with [Fe II] 1.64 $\mu$m emission. However only IRS staring mode is used to observe W49B, the comparison will be possible till new data are obtained from IRS mapping mode observation. In addition, our electron temperature and density estimate for W49B are similar with SNR G349.7+0.2 (7000K and 700 cm$^{-3}$, table 9 of \citealt{hew09}). The molecular temperature estimation is also typical (1000-2000 K for hot component, table 3 of \citealt{neu07}, table 5 of \citealt{hew09}).\\

\begin{figure}[!htpb]

\centerline{\includegraphics[width=0.45\textwidth, angle=0]{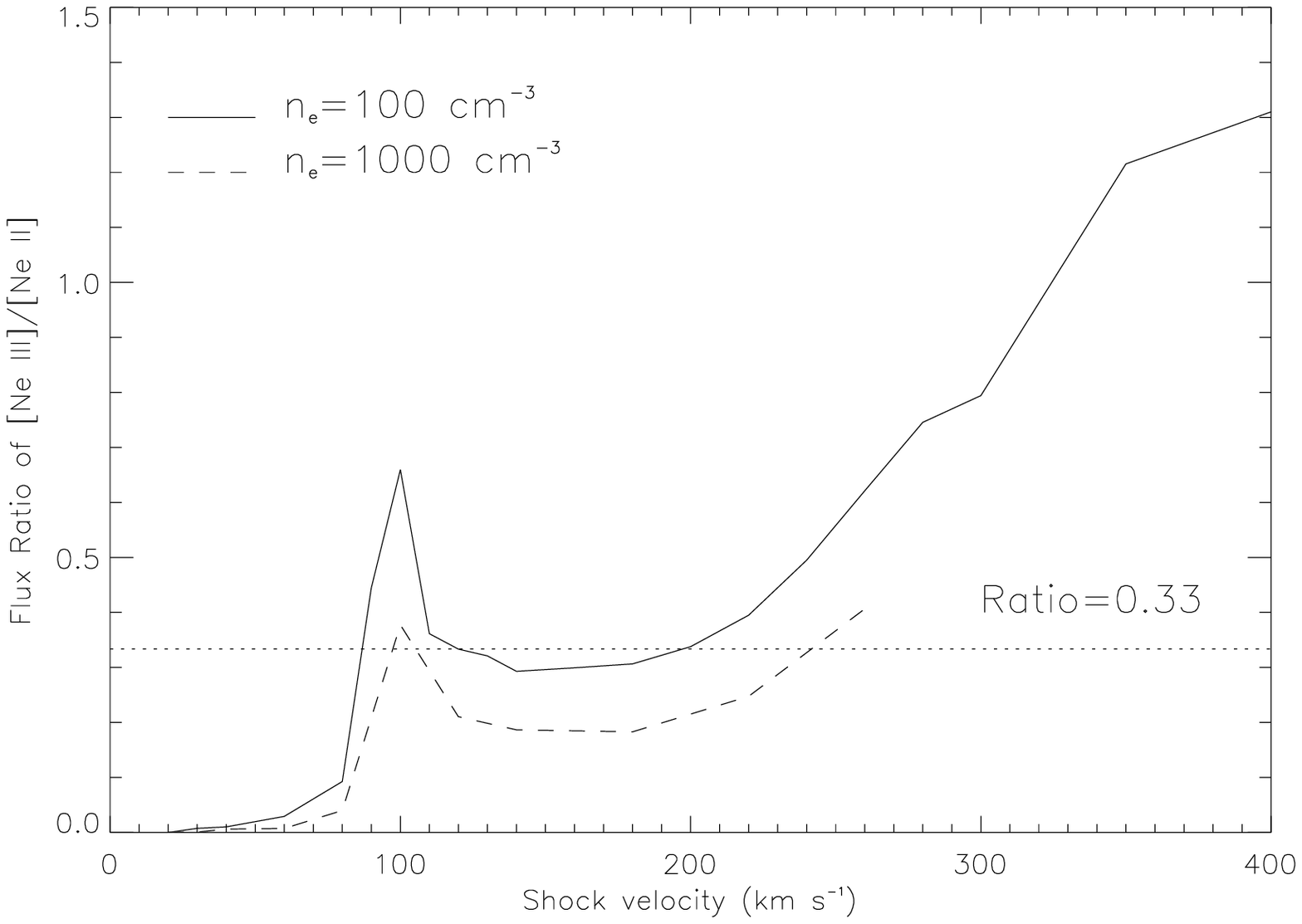}}
\caption{Ratio of [Ne\, III] 15.5 $\mu$m and [Ne\, II] 12.8$\mu$m as a function of shock velocity in the shock model of \citet{har87}. The dotted line represent the ratio.}

\label{fig:13}
\end{figure}

\citet*{hol89} calculated the low excitation lines' intensity in a J-shock with per-shock medium density of ${10}^{3-6}$ cm${}^{-3}$. For W49B, we find a shock with velocity from 60 to 100 km s${}^{-1}$ in ${10}^{3-4}$ cm ${}^{-3} $medium can reproduce the main feature of the measured ionic lines. The [Ne III] to [Ne II] line ratio is another indicator of J-shock velocity (\citealt{and11}, \citealt{hew09}). In Figure \ref{fig:13}, we shows the ratio in the W49B based on the model (\citealt*{har87}), so we suggest a shock velocity of $\sim$90 km s${}^{-1}$.\\

\subsection{The dust}

\subsubsection{A comparison with previous result}

\noindent The dust property of W49B has been studied by \citet*{sak92} (using {\it IRAS} data) and \citet{pin11} (using {\it Spitzer} data). We find obvious discrepancies between the previous results and ours. The inconsistency is likely caused by the low resolution of {\it IRAS} data, the neglected spectral line contribution to the total flux in previous study, and different wavelength coverage between the previous and ours.\\

Using {\it IRAS} data, \citet*{sak92} found W49B has two dust components with temperatures of 157, 35.6 K and dust masses of 0.037 ${\Msol}$, 500 ${\Msol}$ respectively. However, two factors may lead to wrong result when dealing with {\it IRAS} data. The first one is related with {\it IRAS}'s low spacial resolution. As seen in Figure \ref{fig:4}, young stellar object G43.31-0.21 (in the circle) with strong infrared emission is nearby W49B. It is impossible to separate them due to {\it IRAS} low spatial resolution, so the measured flux was the sum of W49B and G43.31-0.21. The second one is that \citet*{sak92} did not consider the contribution of various spectral lines to the total flux, especially in the mid-infrared, i.e. the 12 $\mu$m band of {\it IRAS}. Without the correction for the spectral line contribution, the mass of the hot component was overestimated.\\

\citet{pin11} estimated the dust temperature and mass of W49B are 54 K and 1.6 ${\Msol}$ using {\it Spitzer}'s 24 and 70 $\mu$m data. The weaknesses are from not considering the spectral line contribution at 24 $\mu$m and assuming the flux at 24 $\mu$m and 70 $\mu$m totally comes from the same dust component. Actually, as displayed in Figure \ref{fig:7}, the hot dust contributes the same flux as the warm dust at 24 $\mu$m.\\

Our result also has its weakness. As discussed above, spectral line contribution to the total flux is obvious in the mid-infrared waveband. We try to correct the contribution by the IRS staring mode observation. However, the {\it Spitzer} IRAC images of W49B (\citealt{rea06}) show different type of line emissions at different position of W49B, e.g. more ionic line emissions in the loops and more molecular line emissions at the east and west part. {\it Spitzer} IRS observation only covers a small part of W49B which is not enough for detailed line-contribution correction to the remnant.\\

\subsubsection{The origin of dust}

\noindent Supernova (SN) explosion has been considered as the potential source to produce large amounts of dust in a short timescale, i.e. a few Myr. 2-4 ${\Msol}$ newly formed cold dust has been detected in the young Galactic SNR, Cas A, which has an age of 320 yr (\citealt{dun03}). For W49B, the observed infrared emission is not all from newly formed dust, because the total mass of the ejecta is only about 6 ${\Msol}$ (\citealt{mic08}). We have derived that the total mass of dust associated with W49B is 6.4 $\pm$ 3.2 ${\Msol}$, any newly formed one needs consume all the ejecta to produce dust. Other non-SN origin needs to be explored in order to explain the strong infrared emission of W49B.\\

W49B's infrared morphology follows the radio emission well, the hot component could be explained naturally by the swept up circumstellar and interstellar material. Assuming a gas to dust ratio of 125 (e.g. \citealt*{dra03}), the needed mass of swept up material is a few $\sim$10${}^{-2} {\Msol}$. However, this explanation will not work well for the warm component, because the swept up material will be $\sim$800 ${\Msol}$ which is not consistent with the young age of W49B.\\

\citet{zho11} found the interactions between W49B and the molecular cloud have significant affection on the evolution of the SNR, e.g. the formation of overionized plasma due to the evaporation of the cloud. The work of \citet{ino12} also showed that the densest cloud clumps can survive from the shock and remain inside the SNR. Both works indicate the evaporation of dust from molecular clouds might have a big contribution to the far-infrared emission. Following the work of \citet{lee11}, the expected infrared emission luminosity of an evaporating cloud in the hot gas of an SNRs can be estimated by equation (\citealt*{dwe81}):\\
\begin{equation}
{L}_{IR} \approx 200n_h^2{(\frac{{{R_c}}}{{1pc}})^3}{(\frac{{{T_h}}}{{{{10}^7}K}})^{1.5}}{L_ \odot },
\end{equation}
where ${R_c}$ is the radius of the cloud, ${n_h}$ and ${T_h}$ are the density and temperature of the hot gas. For W49B, ${n_h}$ is about 5 cm${}^{-3}$ (\citealt{mic08}). ${T_h}$ is range from 1.13 keV to 3.68 keV or 1.31 $\times {10^7}$ K to 4.27 $\times {10^7}$ K obtained from X-ray spectral fitting of different regions of W49B and by different authors (\citealt{mic06}, \citealt{keo07}, \citealt{oza09}, \citealt{lop09}, \citealt{mic10}, \citealt{lop13a}). We also estimated the ${T_h}$ by an indirect way using equation (\citealt*{mck80}):\\
\begin{equation}
{T_h} = {(\frac{{{v_r}}}{{839km{s^{ - 1}}}})^2} \times {10^7}K,
\end{equation}
where ${v_r}$ is the forward shock velocity in the plasma. \citet{keo07} estimated the ${v_r}$ to be $\sim$1200 km s${}^{-1}$ for W49B. So the ${T_h}$ is derived to be $\sim2.0 \times {10^7}$ K which is well consistent with the value from X-ray spectral analysis. If we use the value of $2.0 \times {10^7}$ K and assume a radius of $2.3_{ - 0.5}^{ + 0.3}$ pc for the evaporation cloud, the emission of evaporated dusts will have the same luminosity as the observation, i.e. 17.0 $\pm$ 8.5 $\times {10^4} {L_\odot}$ derived from equation, $L = \int {{\kappa _v}B(v,T){M_{dust}}dv}$. Higher ${T_h}$, such as $2.5 \times {10^7}$ K or $3.0 \times {10^7}$ K, will reduce the radius to $2.0_{ - 0.4}^{ + 0.3}$ pc or $1.9_{ - 0.4}^{ + 0.2}$ pc. W49B has a dense environment, dust from evaporation clouds is feasible way to explain the observed far-infrared emission.\\

\section{Summary}

\noindent We study W49B and its environment by radio and infrared observations. We detect fluctuation in the absorption spectrum toward different part of W49B and W49A which support \citet*{bro01}'s suggestion that the difference of {\HI} distribution could explain the absorption difference at $\sim$55 km s${}^{-1}$ between W49B and W49A. We conclude that the previous claim of W49B interacting with clouds at the southern boundary at velocity of $\sim$5 km s${}^{-1}$ is likely not right. W49B is instead of likely associated with molecular cloud at $\sim$40 km s${}^{-1}$, this suggests that W49B has a distance of $\sim$10 kpc.\\

{\it Spitzer} IRS observation reveals clear pure rotational shocked H$_2$ lines, H$_2$ (0,0) S(0)-S(7), which supports W49B is interacting with molecular cloud. Boltzmann diagram suggests there are two components of H$_2$ with temperatures of $\sim$260 K and $\sim$1060 K. Spectral lines of S$^0$, S$^{2+}$, Ar$^{+}$, Cl$^{+}$, Fe$^{+}$, Ne$^{+}$, Ne$^{2+}$ and Si$^{+}$ are also detected. We find the ionic phase has electron density of $\sim$500 cm${}^{-3}$ with a temperature of $\sim{10^4}$ K. A J-shock with velocity of $\sim$90 km s${}^{-1}$ may produce the main ionic spectral line features.\\

Mid- and far-infrared SED fitting implies there are two components of dust with temperature of 151 $\pm$ 20 K and 45 $\pm$ 4 K associated with W49B. Each components have masses of 7.5 $\pm$ 6.6 $\times$ ${10}^{-4} {\Msol}$ and 6.4 $\pm$ 3.2 ${\Msol}$ respectively. The hot dust can be explained by the swept up circumstellar or interstellar materials. The warm dust may originate from evaporation of the clouds interacting with W49B.\\

\begin{acknowledgements}
HZ and WWT acknowledge supports from NSFC (211381001, Y211582001) and BaiRen programme of the CAS (034031001). This work is partly supported by China's Ministry of Science and Technology under State Key Development Program for Basic Research (2012CB821800, 2013CB837901). We thank Drs Y. Zhang, X.L. Liu, P. Wei, X.M. Wang and Y. Su for meaningful discussion when preparing this paper. This research has made use of the NASA/ IPAC Infrared Science Archive, which is operated by the Jet Propulsion Laboratory, California Institute of Technology, under contract with the National Aeronautics and Space Administration.\\
\end{acknowledgements}

\bibliographystyle{apj}
\bibliography{bibtex}

\end{document}